\documentclass[12pt]{article}
\usepackage{amsmath}
\usepackage{graphicx}
\usepackage{enumerate}
\usepackage{natbib}
\usepackage{url} 

\usepackage{amsfonts}
\usepackage[pdftex, colorlinks, urlcolor=blue, linkcolor=blue, anchorcolor=blue, citecolor=blue]{hyperref}

\usepackage{amsfonts}
\usepackage{bm}
\usepackage{indentfirst}
\usepackage{float}
\usepackage{setspace}
\usepackage{textcomp}
\usepackage{makecell}
\usepackage{multirow}
\usepackage{amssymb}
\usepackage{color}
\usepackage{subcaption}

\newcommand{\blind}{1}

\begin{document}

\def\spacingset#1{\renewcommand{\baselinestretch}%
{#1}\small\normalsize} \spacingset{1}


\if1\blind
{
  \title{\bf Convolutional Non-homogeneous Poisson Process with Application to Wildfire Risk Quantification for Power Delivery Networks}
  \author{Guanzhou Wei\\
    Department of Industrial Engineering, University of Arkansas\\
    Feng Qiu \\
    Argonne National Lab\\
    Xiao Liu \\
    Department of Industrial Engineering, University of Arkansas
    }
  \maketitle
} \fi

\if0\blind
{
  \bigskip
  \bigskip
  \bigskip
  \begin{center}
    {\LARGE\bf Convolutional Non-homogeneous Poisson Process with Application to Wildfire Risk Quantification for Power Delivery Networks}
\end{center}
  \medskip
} \fi

\bigskip
\begin{abstract}
The current projection shows that much of the continental U.S. will have significantly hotter and drier days in the following decades, leading to more wildfire hazards that threaten the safety of power grid. Unfortunately, the U.S. power industry is not well prepared and still predominantly relies on empirical fire indices which do not consider the full spectrum of dynamic environmental factors. This paper proposes a new spatio-temporal point process model, Convolutional Non-homogeneous Poisson Process (cNHPP), to quantify wildfire risks for power delivery networks. The proposed model captures both the current short-term and cumulative long-term effects of covariates on wildfire risks, and the spatio-temporal dependency among different segments of the power delivery network. The computation and interpretation of the intensity function are thoroughly investigated, and the connection between cNHPP and Recurrent Neural Network is also discussed. We apply the proposed approach to estimate wildfire risks on major transmission lines in California, utilizing historical fire data, meteorological and vegetation data obtained from the National Oceanic and Atmospheric Administration and National Aeronautics and Space Administration. Comparison studies are performed to show the applicability and predictive capability of the proposed approach. Useful insights are obtained that potentially enhance power grid resilience against wildfires. 
\end{abstract}

\noindent%
\vfill

\spacingset{1.9} 
\section{Introduction} \label{sec:intro}
\vspace{-8pt}
Wildfires, ignited naturally (e.g., lightning) or by power delivery, are increasingly threatening energy infrastructure and public safety, and sometimes evolve into disastrous events. During the past U.S. wildfire events, an average of 3.3 million acres were burned annually in the 1990s, while this figure has been more than doubled to 7.0 million acres since 2000 \citep{hoover2021wildfire}.
The current projection shows that much of the continental U.S. will have significantly hotter and drier days due to climate change \citep{brown2021us}, leading to more wildfire hazards threatening the safety and reliability of the power grid.

However, the U.S. power industry and utilities are not well prepared and still predominantly rely on empirically calculated fire indices for wildfire risk analysis \citep{SDGE}. These indices are calculated from a set of environmental variables using some predefined and relatively simple formulas, and therefore, do not consider the full spectrum of dynamic environmental factors such as real-time meteorological variables.   
Hence, understanding and comprehensively quantifying wildfire risks for power delivery infrastructures, given meteorological and vegetation variables, become critical for enhancing power systems resilience. In this paper, we propose a new spatio-temporal point process model known as the Convolutional Non-homogeneous Poisson Process (\texttt{cNHPP}), and apply the approach to quantify wildfire risks for major power transmission lines in California.

In 2018, the Camp Fire (known as the deadliest wildfire in California history) was ignited by a faulty electric transmission line. The Camp Fire killed 85 people, destroyed  over 15,000 structures, and caused a total insured loss of 12.5 billion \citep{schulze2020wildfire}. This is only a snapshot of the significant impact of increasingly frequent wildfires on power delivery infrastructures and communities. 
The Thomas Fire in 2017 interrupted the power transmission lines in the Santa Barbara area, and 85,000 customers lost their electric power \citep{nazaripouya2020power}.  
Several other worst wildfires were also ignited by power lines including the Grass Valley Fire, Malibu Canyon Fire, Rice Fire, Sedgwick Fire and, Witch Fire, and these power-line fires burned a total area of more than 334 square miles \citep{CPUC}. 

In response to the devastation of power-line fires, the California Public Utilities Commission (CPUC) launched the Public Safety Power Shutoff (PSPS) activity in 2018 which authorizes electric utilities to shut off electric power for public safety. Pacific Gas and Electric (PG\&E), one of the major utilities in California, also took a series of PSPS activities in 2019. Although these PSPS activities decreased the ignition of power-line fires, about 1,848,000 customers were impacted by a series of power outages and 44\% of respondents reported power loss for three or more days \citep{mildenberger2022effect}.

\begin{figure}[h!]
    \centering
    \includegraphics[width=0.45\textwidth]{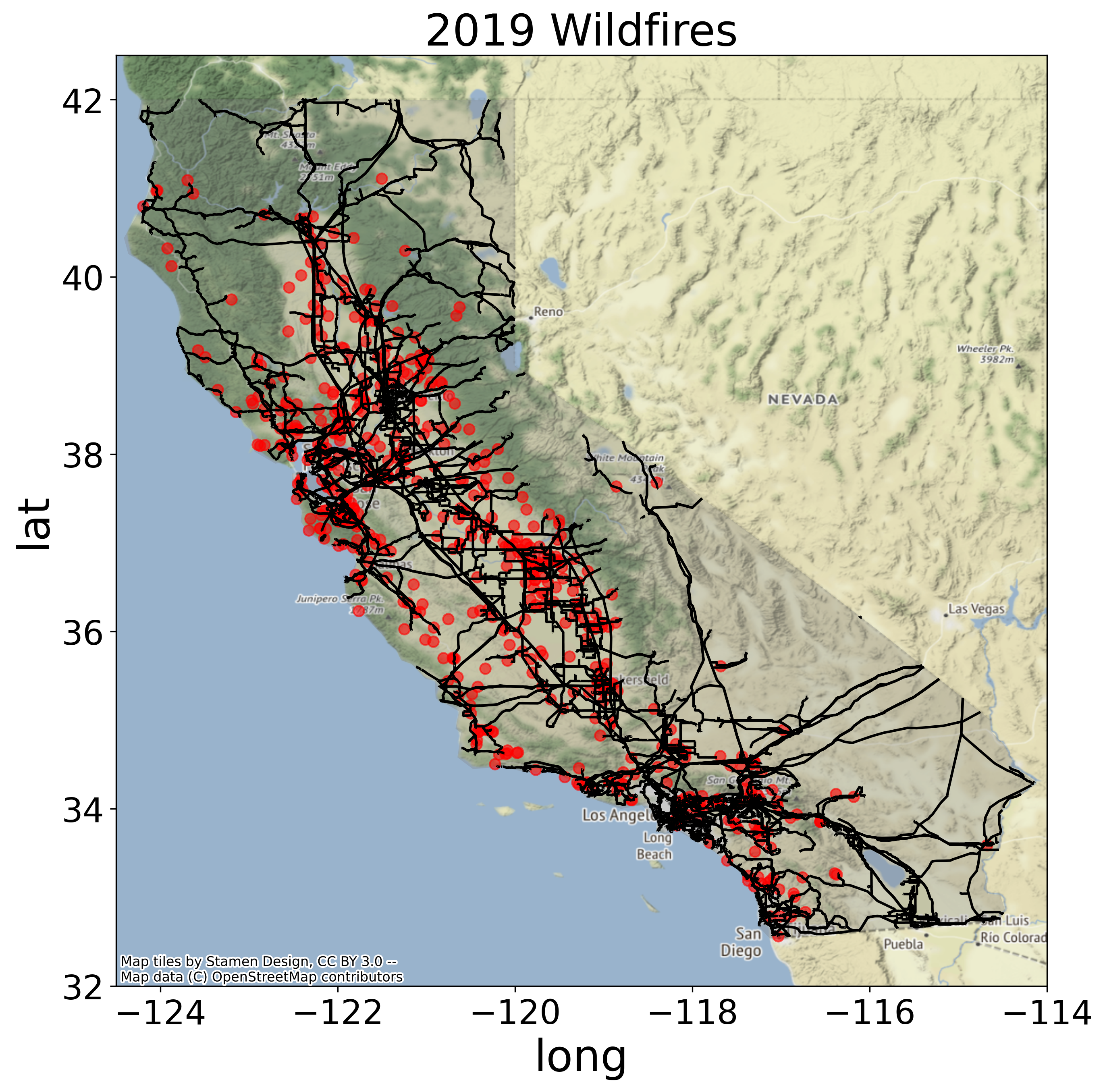}
    \caption{Wildfire incidents from California Public Utilities Commission and major networks of power transmission lines in California in 2019.}
    \label{fig:Motivating_example}
\end{figure}

\vspace{-8pt}
To better plan for PSPS against wildfires,  prioritize power grids protection measures, and utilize limited resources, accurate wildfire risk quantification and prediction are often needed at different segments/locations of power grids \citep{rhodes2022co}. 
This motivates us to propose a spatio-temporal point process model that can be applied to quantify wildfire risks on networks of power transmission lines. Ideally, the model needs to be capable of capturing (i) (short-term) current effects of environmental covariates on fire risks, (ii) (long-term) cumulative effects of historical covariates information on current fire risks, and (iii) spatio-temporal interactions among neighboring power transmission lines. This objective is made possible as data become readily available. For example, the CPUC requires electric companies to report fire incident data from their power facilities. Figure \ref{fig:Motivating_example} shows the reported power-line fires in 2019 and the network of power transmission lines in California from the U.S. Energy Information Administration (EIA). It can be seen that most of the fire incidences are in close proximity to power transmission lines and are closely related to power delivery infrastructures. Together with the meteorological and vegetation variables from the National Oceanic and Atmospheric Administration (NOAA) and National Aeronautics and Space Administration (NASA), we have the data needed for constructing and validating the statistical model to be described in this paper that estimates and predicts wildfire event intensities on segments of a network of power transmission lines. 

\vspace{-8pt}
\subsection{Literature Review}
\vspace{-8pt}
Assessment of wildfire hazards has a long history since the Canadian Forest Fire Danger Rating System (CFFDRS) was developed in 1968 and the United States National Fire Danger Rating System (NFDRS) was created in 1972. These rating systems generate fire indices that reflect the potential wildfire hazards based on weather, fuels, and topography information \citep{NFDRS}; for example, the Canadian Fire Weather Index (CFWI) and Fire Potential Index (FPI). These indices are often directly calculated based on the physical knowledge about wildfires and environmental processes, and are used to capture the overall long-term trend of wildfire risks on a large scale. Hence, these fire indices may not be used to quantify wildfire event intensities for local regions within short time windows. Given observed wildfire incident data, point process models have been widely used to quantify wildfire risks \citep{taylor2013wildfire,holbrook2022bayesian}. 
For example, \citet{peng2005space} developed a spatio-temporal point process for modeling wildfire risks in Los Angeles County. The authors incorporated the burning index (one of the indices generated from the NFDRS), as well as the space and time trends obtained from the historical wildfire data into the intensity function of the proposed spatio-temporal point process. \citet{xu2011point} found that the incorporation of weather information and fuel age into the intensity function largely enhanced the performance of the spatio-temporal point process proposed in \citet{peng2005space} for modeling wildfires in Los Angeles County. \citet{serra2014spatio} adopted the spatio-temporal log-Gaussian Cox processes for modeling Catalonia wildfire occurrences. \citet{opitz2020point} leveraged a log-Gaussian Cox process for forest fires in Mediterranean France, which incorporates covariates such as land use and weather conditions through a linear model with random effects. 

Although various spatio-temporal point processes have been investigated for modeling wildfire risks, most of these approaches primarily model wildfire risks in a continuous two-dimensional space. Because power delivery infrastructures are distributed on a linear network that consists of segments of power lines, there exists a need to construct spatio-temporal point processes on a linear network of power transmission lines, implying that the support of the process is constrained by a linear network. \citet{uppala2020modeling} adopted the separable temporal linear point process to model wildfire ignition on a road network, in which the Papangelou conditional intensity takes a log-linear form of the covariates. \citet{zhu2022spatio} used the Hawkes process to model power outages on the power grid under extreme weather. The authors constructed the background intensity using a deep neural network with temporally cumulative weather effects, and the triggering effects were formulated by the power grid connectivity and outage history.  Note that, point process models on linear networks have also been found in the modeling of street crimes and traffic accidents \citep{baddeley2021analysing},
visitors’ stops at touristic attractions \citep{d2022inhomogeneous}, ambulance interventions on a road network \citep{gilardi2021non}, etc. 

\vspace{-8pt}
\subsection{Overview and Contributions}
\vspace{-8pt}
Based on the discussions above, this paper proposes a spatio-temporal point process model on a linear network and applies the model to estimate fire event intensities on segments of power transmission lines. The contributions are summarized as follows:

$\bullet$ We propose a new spatio-temporal point process model, known as the Convolutional Non-homogeneous Poisson Process (\texttt{cNHPP}), on a linear network. Based on the proposed model, the event process on each segment of a linear network is modeled as an NHPP with its log intensity being given by an infinite series. For each segment $i$ of the network, the model captures (i) how the current covariates associated with segment $i$ affect the event intensity on this segment (i.e., the short-term instantaneous effects), and (ii) how the historical covariates associated with segment $i$ and its neighboring segments together affect the event intensity of segment $i$ (i.e., the long-term cumulative effects) given network topology and spatio-temporal dependency among segments. In particular, the current effects of covariates on event intensity are modeled through a log-linear model, while the cumulative historical effects are captured by a convolution approach. Note that, the proposed \texttt{cNHPP} is different from the self-exciting spatio-temporal point process, which describes the intensity function as the sum of the current effects and the effects from historical events in space and time \citep{mohler2011self, holbrook2022bayesian}.
For the self-exciting process, the historical effects are only triggered by the occurrences of historical events. This may not always be effective in modeling wildfire events which are considered as rare events (in the example presented in Section 3, only 15 fire events are recorded within a month from a network of 7000 power transmission lines). In contrast, the convolution approach adopted in this paper establishes the spatial interactions and temporal dependence by modeling how historical covariate information from neighboring segments affect the intensity of a given segment, regardless of whether there are any historical fire events (In other words, instead of letting historical events trigger the spatio-temporal interactions, the proposed model directly uses historical covariate information over the entire network to establish the spatio-temporal dependency). In such a way, the cumulative historical and spatial effects play a much stronger role in the proposed model; see Section 2.1. 

$\bullet$ In Section 2.2 and 2.3, we provide additional insights and discussions on the proposed  \texttt{cNHPP}. In particular, we present detailed investigations on computing the proposed intensity function, as well as the graphical representation of the proposed model. Furthermore, we draw the connection between the proposed  \texttt{cNHPP} and Recurrent Neural Network (RNN). 

$\bullet$ In Section 3, utilizing the environmental data obtained from NOAA and NASA, we  apply the proposed approach to model and predict the wildfire risks on major transmission lines in parts of California. By investigating the estimated effects of different covariates on the occurrences of wildfires on transmission lines, we obtain useful insights and recommendations for power system operation, protection and maintenance that potentially enhance power grid resilience against wildfires. Section 4 concludes the paper. 

\vspace{-8pt}
\section{Convolutional NHPP on a Linear Network}
\vspace{-8pt}
In Section \ref{sec:model_formulation}, we first present the proposed convolutional NHPP model. Section \ref{sec:graphic_represetation} presents the graphical representation of the proposed model and performs in-depth investigations on the computation of the intensity function. In Section \ref{sec:model_inherited_rnn}, we provide a different perspective of the proposed model by drawing its connection to RNN. 

\vspace{-8pt}
\subsection{Basic Model Formulation}\label{sec:model_formulation}
\vspace{-8pt}
Consider a linear network $L$ with $N$ segments; e.g., power transmission lines. In particular, let $l_i$ be the $i$-th segment, and the  network can be represented by $L=\cup_{i=1}^{N}l_i$. In this work, we consider events that occur on segments of a network, and the event process on the $i$-th segment is modeled as a point process with a conditional intensity function:
\vspace{-8pt}
\begin{equation}
    \lambda(i, t|\mathcal{H}_t) = \lim_{\Delta\rightarrow0} \frac{\mathbb{E}[N_i[t, t+\Delta)|\mathcal{H}_t]}{\Delta},
\end{equation}

\vspace{-8pt}
\noindent where $N_i[t, t+\Delta)$ is a counting measure on the $i$-th segment on the time interval $[t, t+\Delta t)$, and $\mathcal{H}_t$ represents the event history on the entire network $L$ (which is omitted thereafter). In particular, we assume that the event process on each segment  $i$ is an NHPP with the intensity function $\lambda(i, t)$. Hence, for any $\tilde{L} \subseteq  L$, the total number of events on the sub-network $\tilde{L}$ at a time interval $[t,t+\Delta)$ follows a Poisson distribution with parameter $\int_{t}^{t+\Delta}\sum_{i \in \tilde{L}}\lambda(i, u)du$. Since segments $l_i$, $i=1,2,\cdots,N$, in a network are disjoint, we have 
\vspace{-8pt}
\begin{equation} \label{eq:event_prob}
    \text{Pr}(N_i[t, t+\Delta)=n_i, i=1,2,\cdots,N) = \prod_{i}^{N}\frac{ \Lambda^{n_i} }{n_i!}e^{-\Lambda}
\end{equation}

\vspace{-8pt}
\noindent where $n_i$ is a non-negative integer and $\Lambda = \int_{t}^{t+\Delta}\lambda(i, u)du$. For power line fire risk quantification, for example, the equation above allows us to evaluate the probabilities of different fire scenarios over a network. 

Hence, the construction of the intensity function $\lambda(i, t)$ becomes critical. The goal of this paper is to construct a statistical model that adequately explains the intensity function $\lambda(i, t)$ by taking into account both the short-term (i.e., current) and long-term (i.e., cumulative) effects of covariates, as well as the spatio-temporal interactions among multiple segments. The following model is proposed,
\vspace{-12pt}
\begin{equation}\label{eq:intensity}
    \log\lambda(i, t) = \underbrace{c(i,t)}_{\text{current effects}} + \underbrace{h(i,t)}_{\substack{\text{cumulative historical}\\\text{and spatial effects}}}
\end{equation}

\vspace{-8pt}
\noindent which decomposes the (log) intensity into two additive components. The first component $c(i,t)$ captures the current effects of covariates at time $t$, while the second component $h(i,t)$ captures the cumulative effects of historical covariate information (before time $t$) through the spatio-temporal interactions among neighboring network segments. To elaborate, 

$\bullet$ $c(i,t)$ incorporates the current effect of covariates at time $t$ for segment $i$ through a linear model $c(i,t)=\bm{x}^T(i,t)\bm{\beta}$, where $\bm{x}(i,t) = (1, x_1(i,t), x_2(i,t), \cdots, x_q(i,t))^T$ denote the  covariates associated with segment $i$ at time $t$, $\bm{\beta}=(\beta_0, \beta_1, \cdots, \beta_q)^T$ is a vector of covariate effects, and $q$ is the number of covariates. In the application presented in Section \ref{sec:application}, potential covariates for fire events include vegetation and meteorological variables. 

$\bullet$ $h(i,t)$ captures the long-term cumulative effects as well as the spatio-temporal interactions among segments of a network. In other words, it explains how historical covariate information (before time $t$) associated with the neighboring segments of segment $i$ affects the intensity of segment $i$ at time $t$. We model $h(i,t)$ in the following way:
\vspace{-12pt}
\begin{equation}\label{eq:convolution}
    h(i,t) = \xi\sum_{i'\in\Omega_i}w_{ii'}\log\lambda(i', t-\Delta),
\end{equation}

\vspace{-8pt}
\noindent where $\Omega_i$ is the set that contains  pre-defined neighboring segments of the $i$-th segment, 
$w_{ii'}$ is the contribution (i.e., weight) to $h(i,t)$ from the $i'$th segment from time $t-\Delta$, and $\xi \in [0,1)$ is the decay factor that controls the rate of decay of the cumulative effects. Note that, a smaller $\xi$ indicates that the spatial cumulative effect quickly decays, while a larger $\xi$ makes the current intensity to be dependent more on historical intensities. In other words, a larger $\xi$ makes $\lambda(i,t)$ smoother in time. This idea is similar to  the Exponentially-Weighted Moving Average  that controls the smoothness of the moving averages by distributing weight to current and historical observations. In an extreme case when $\xi=0$, the proposed approach degenerates to a conventional NHPP model with a log-linear intensity function. 

In fact, the model (\ref{eq:convolution}) implies that the intensity at time $t$ depends not only on the intensity at time $t-\Delta$, but also on  the intensities at times $t-2\Delta, t-3\Delta, \cdots$. One may see this by replacing $\log \lambda(i', t-\Delta)$ in (\ref{eq:convolution}) by $c(i', t-\Delta)+h(i',t-\Delta)$. By iterating  this process, the intensity of each segment depends on that of neighboring segments over the entire history.
To elaborate how the spatio-temporal dependency structure among $\{\lambda(i,t)\}_{i=1}^{N}$ is established by (\ref{eq:convolution}), for a function $f(i,t)\colon\{1,2,\cdots,N\}\times[0,T]\mapsto \mathbb{R}^+$, we introduce a \textit{Network Convolution} operator $\mathcal{NC}$:
\vspace{-12pt}
\begin{equation}\label{eq:operator}
    \mathcal{NC}\{f\}(i,t) = \sum_{i'\in\Omega_i}w_{ii'}f(i',t),
\end{equation}

\vspace{-8pt}
\noindent where $\mathcal{NC}\{f\}(i,t)\colon\{1,2,\cdots,N\}\times[0,T]\mapsto \mathbb{R}^+$ is a new function generated by the operator $\mathcal{NC}$. Note that, $\mathcal{NC}$ is not strictly a mathematical convolution operation but a linear combination functions, similar to the interpretation of convolution in a Convolutional Neural Network (CNN). As to be discussed in Section \ref{sec:graphic_represetation}, the $\mathcal{NC}$ operator defined in (\ref{eq:operator}) imposes sparsity in the weight matrix $\bm{W}$. Because the operator $\mathcal{NC}$ is linear, we may also define $\mathcal{NC}^{(n)}\{f\}(i,t)$ as the $n$-fold network convolution for a function $f(i,t)$. For example, applying $\mathcal{NC}$ to $f(i,t)$ twice (i.e., a two-fold operation) yields:
\vspace{-8pt}
\begin{equation}
    \begin{aligned}
            \mathcal{NC}^{(2)}\{f\}(i,t) & = \mathcal{NC}\{\mathcal{NC}\{f\}\}(i,t) = \mathcal{NC}\left\{\sum_{i'\in\Omega_i}w_{ii'}f(i',t)\right\}\\
                               & = \sum_{i'\in\Omega_i}w_{ii'}\mathcal{NC}\{f\}(i',t) = \sum_{i'\in\Omega_i}w_{ii'}\sum_{i''\in\Omega_{i'}}w_{i'i''}f(i'',t).
    \end{aligned}
\end{equation}

Based on the $\mathcal{NC}$ operator (\ref{eq:operator}) and the model (\ref{eq:intensity}), $h(i,t)$ in (\ref{eq:convolution}) can be written as $ h(i,t)=\sum_{n=1}^{\infty}\xi^n\mathcal{NC}^{(n)}\{c\}(i,t-n\Delta)$. The derivation of $h(i,t)$ involves a long equation that is provided in the Supplementary Material. 
Finally, by substituting the expression of $ h(i,t)$ into (\ref{eq:intensity}), we obtain the proposed statistical model as follows: 

\textbf{Convolutional NHPP.} \textit{Consider a linear network $L=\cup_{i=1}^{N}l_i$ with $N$ segments. The  event process on each segment $i$ is modeled as an NHPP with its log intensity being given by an infinite series:}
 \vspace{-14pt}
\begin{equation}\label{eq:analyticform}
    \begin{aligned}
        \log \lambda(i,t) &= c(i,t) + h(i,t) = \mathcal{NC}^{(0)}\{c\}(i,t) + \sum_{n=1}^{\infty}\xi^n\mathcal{NC}^{(n)}\{c\}(i,t-n\Delta) \\
                          &= \sum_{n=0}^{\infty}\xi^n\mathcal{NC}^{(n)}\{c\}(i,t-n\Delta),
    \end{aligned}
\end{equation}

 \vspace{-14pt}
\noindent \textit{where $\mathcal{NC}^{(0)}\{c\}(i,t) \triangleq c(i,t)$}.

It is clearly seen that, 

(i) The intensity of segment $i$ at time $t$ depends on not only the covariate information associated with segment $i$ at time $t$ (i.e., the current effect), but also the historical covariate information associated with neighboring segments prior to time $t$ (i.e., the cumulative historical and spatio-temporal dependency).   

(ii) The (log) intensity $\log \lambda(i,t)$ is represented by the sum of an infinite series. Each term $\xi^n\mathcal{NC}^{(n)}\{c\}(i,t-n\Delta)$ corresponds to the contribution to $\log \lambda(i,t)$ from $c(\cdot, t-n\Delta)$ at the current or a historical time $t-n\Delta$, $n=0,1,2,\cdots$. Because the contribution from $t-n\Delta$ decays to zero as $n$ increases for $\xi \in [0,1)$, it is possible to truncate the series by only retaining the first $K$ terms. In the next section, we present a graphical representation of the model and illustrate the computation of $\log \lambda(i,t)$ leveraging such a graphical representation.    

(iii) The proposed model is different from the well-known self-exciting spatio-temporal point process, 
such as the Hawkes Process \citep{mohler2011self},
which describes the intensity function as the sum of current effects and the effects from historical events:
 \vspace{-14pt}
\begin{equation}\label{eq:intensity_self}
    \log\lambda(i, t) =c(i,t)+ \sum_{j:t_j<t} g(i,s_j, t, t_j),
\end{equation}

 \vspace{-14pt}
\noindent where $g$ is known as the triggering function, $s_1, s_2, \cdots, s_r$ are the segments where historical events occur, $t_1, t_2, \cdots, t_r$ are the times when historical events occur, and $r$ is the total number of historical events. Following the self-exciting process, whenever an event occurs at a spatial location in the past, the occurrence of that event affects the current intensity function over the spatial domain through a non-negative triggering function $g$ (e.g., a kernel function or power law decay function). In other words, the historical effects are only triggered by the occurrences of historical events. This may not always be effective in modeling rare events which are sparse in space and time; such as wildfires. For example, in the example presented in Section \ref{sec:application}, 15 out of 7000 power transmission lines experience fire events within a one-month period. Although 15 events can already significantly impact power grid operation, it is a small number considering the size of the power transmission network. Hence, when a traditional self-exciting point process is adopted, the historical effects (only occur when there is a historical event) may not significantly change the intensity over the entire spatial domain. In contrast, the convolution approach adopted in this paper establishes the spatial interactions and temporal dependence by modeling how historical covariate information from neighboring segments affect the intensity of a given segment, regardless of whether there are any historical fire events. Unlike self-exciting processes for which historical events trigger the spatio-temporal interactions, the proposed model directly uses historical covariate information over the entire network to establish the spatio-temporal dependency. In such a way, the cumulative historical and spatial effects play a much stronger role in the proposed model. 

In addition, the convolution operation captures the spatial and temporal interactions of cumulative effects in a space-time inseparable manner. Although non-separable space-time covariance structures have been investigated in spatio-temporal models \citep{stein2005space,kuusela2018locally,katzfuss2020ensemble}, triggering functions are often chosen to be space-time separable in existing self-exciting process models.

\subsection{Computation, Graphical Representation and Likelihood}\label{sec:graphic_represetation}

\begin{figure}[h!]
    \centering
    \includegraphics[width=0.7\textwidth]{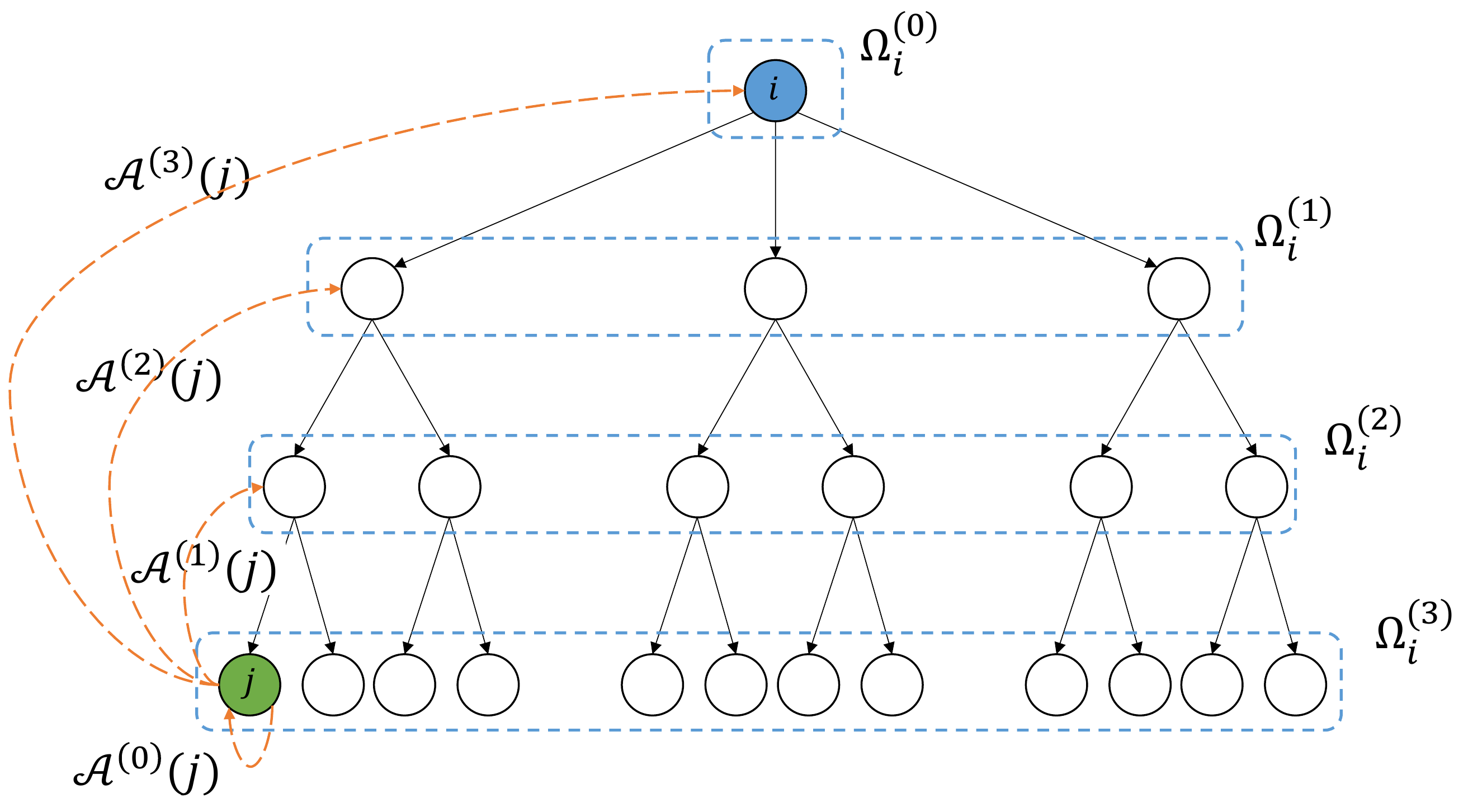}
    \caption{A graphical illustration of neighbor set and ancestor operation with each node representing a segment: (i) $\Omega_i^{(0)}$ is segment $i$ itself, $\Omega_i^{(1)}$ contains the neighbors of segment $i$, $\Omega_i^{(2)}$ contains the neighbors of the neighbors of segment $i$, and so on; (ii) $\mathcal{A}^{(0)}(j)$ return $j$ itself, $\mathcal{A}^{(1)}(j)$ returns the parent of $j$, $\mathcal{A}^{(2)}(j)$ returns the grand parent of $j$, and so on.}
    \label{fig:graph}
\end{figure}

In this subsection, we show how the computation of $\log \lambda(i,t)$ in (\ref{eq:analyticform}) takes a natural graphical representation, and present the likelihood function needed for parameter estimation. We first introduce some notation and network operations. For any segment $i$, let $\Omega_i^{(m)}$ be a set that contains the indices of the $m$-th generation neighbors of segment $i$ ($m=0,1,2,\cdots,K$). As illustrated in Figure \ref{fig:graph}, $\Omega_i^{(0)}=\{i\}$ contains segment $i$ itself, $\Omega_i^{(1)}$ contains the immediate neighbors of segment $i$, $\Omega_i^{(2)}$ contains the neighbors of the neighbors of segment $i$, and so on. 

 \vspace{-2pt}

\vspace{-14pt}

In addition, for any line segment $j$ in the set $\Omega_i^{(m)}$, we also define the \emph{ancestor} operator $\mathcal{A}^{(1)}(j)$ that returns the parent segment of $j$. Similarly, we may introduce the 2-fold ancestor operation $\mathcal{A}^{(2)}(j)$ that returns the grand parent of $j$. By extending this idea, we let $\mathcal{A}^{(n)}(j)$ denote the $n$-fold ancestor operation and let $\mathcal{A}^{(0)}(j)$ return $j$ itself; see Figure \ref{fig:graph}. It is easy to see that, for any line segment $j$ that belongs to the $m$-th generation neighbor set $\Omega_i^{(m)}$ of segment $i$, the $m$-fold ancestor operation of $j$ returns $i$, i.e., $\mathcal{A}^{(m)}(j)=i$ for $j \in \Omega_i^{(m)}$.

Based on the notation and operation defined above, we show that the computation of $\log \lambda(i,t)$ takes a natural graphical representation. 

$\bullet$ (contribution from time $t$) The contribution to $\log\lambda(i,t)$ at time $t$ directly comes from $c(i, t)$, which is the first term of the series (\ref{eq:analyticform}).

$\bullet$ (contribution from time $t-\Delta$) The contribution to $\log\lambda(i,t)$ from $c(\cdot, t-\Delta)$ at time $t-\Delta$ is associated with all neighboring segments in $\Omega_i^{(1)}$, i.e., the second term of the series (\ref{eq:analyticform}) can be computed by
 \vspace{-18pt}
\begin{equation}
    \xi\mathcal{NC}^{(1)}\{c\}(i,t-\Delta) =  \xi\sum_{j\in\Omega_i^{(1)}}w_{\mathcal{A}^{(1)}(j)\mathcal{A}^{(0)}(j)}c(j, t-\Delta).
\end{equation}
 \vspace{-36pt}
 
$\bullet$ (contribution from time $t-2\Delta$) The contribution to $\log\lambda(i,t)$ from $c(\cdot, t-2\Delta)$ at time $t-2\Delta$ is associated with all neighboring segments in $\Omega_i^{(2)}$, i.e., the third term of the series (\ref{eq:analyticform}) can be computed by
 \vspace{-18pt}
\begin{equation}
    \xi^2\mathcal{NC}^{(2)}\{c\}(i,t-2\Delta) =  \xi^2\sum_{j\in\Omega_i^{(2)}}w_{\mathcal{A}^{(2)}(j)\mathcal{A}^{(1)}(j)}w_{\mathcal{A}^{(1)}(j)\mathcal{A}^{(0)}(j)}c(j, t-2\Delta).
\end{equation}
 \vspace{-36pt}
 
$\bullet$ (contribution from time $t-3\Delta$) The contribution to $\log\lambda(i,t)$ from $c(\cdot, t-3\Delta)$ at time $t-3\Delta$ is associated with all neighboring segments in $\Omega_i^{(3)}$, i.e., the fourth term of the series (\ref{eq:analyticform}) can be computed by
 \vspace{-18pt}
\begin{equation}
   \xi^3\mathcal{NC}^{(3)}\{c\}(i,t-3\Delta) =  \xi^3\sum_{j\in\Omega_i^{(2)}}w_{\mathcal{A}^{(3)}(j)\mathcal{A}^{(2)}(j)}w_{\mathcal{A}^{(2)}(j)\mathcal{A}^{(1)}(j)}w_{\mathcal{A}^{(1)}(j)\mathcal{A}^{(0)}(j)}c(j, t-3\Delta).
\end{equation}
 \vspace{-36pt}
 
$\bullet$ (the general case at time $t-k\Delta$) In fact, for any given $m \in \mathbb{N}^+$, it is seen that the generic expression of the contribution to $\log\lambda(i,t)$ from $c(\cdot, t-k\Delta)$ is associated with all neighboring segments in $\Omega_i^{(k)}$ and can be written as
 \vspace{-12pt}
\begin{equation} \label{eq:expression}
    \xi^k\mathcal{NC}^{(k)}\{c\}(i,t-k\Delta) = \xi^k\sum_{j\in\Omega_i^{(k)}}\left\{\prod_{p=1}^{k} w_{\mathcal{A}^{(p)}(j)\mathcal{A}^{(p-1)}(j)}  \right\}c(j, t-k\Delta).
\end{equation}
 \vspace{-28pt}

The generic expression (\ref{eq:expression}) above provides a way to compute each term in the series (\ref{eq:analyticform}). When the   series (\ref{eq:analyticform}) is truncated by only retaining the first $K$ terms, we have 
 \vspace{-14pt}
\begin{equation}\label{eq:totalcontribution}
\begin{split}
    \log\lambda(i,t) & = \sum_{n=0}^{\infty}\xi^n\mathcal{NC}^{(n)}\{c\}(i,t-n\Delta)\\
    & \approx c(i,t) + \sum_{n=1}^{K} 
    \xi^n\sum_{j\in\Omega_i^{(n)}}\left\{\prod_{p=1}^{n} w_{\mathcal{A}^{(p)}(j)\mathcal{A}^{(p-1)}(j)}  \right\}c(j, t-n\Delta).
\end{split}   
\end{equation}
 \vspace{-28pt}
 
\begin{figure}[h!]
    \centering
    \includegraphics[width=1\textwidth]{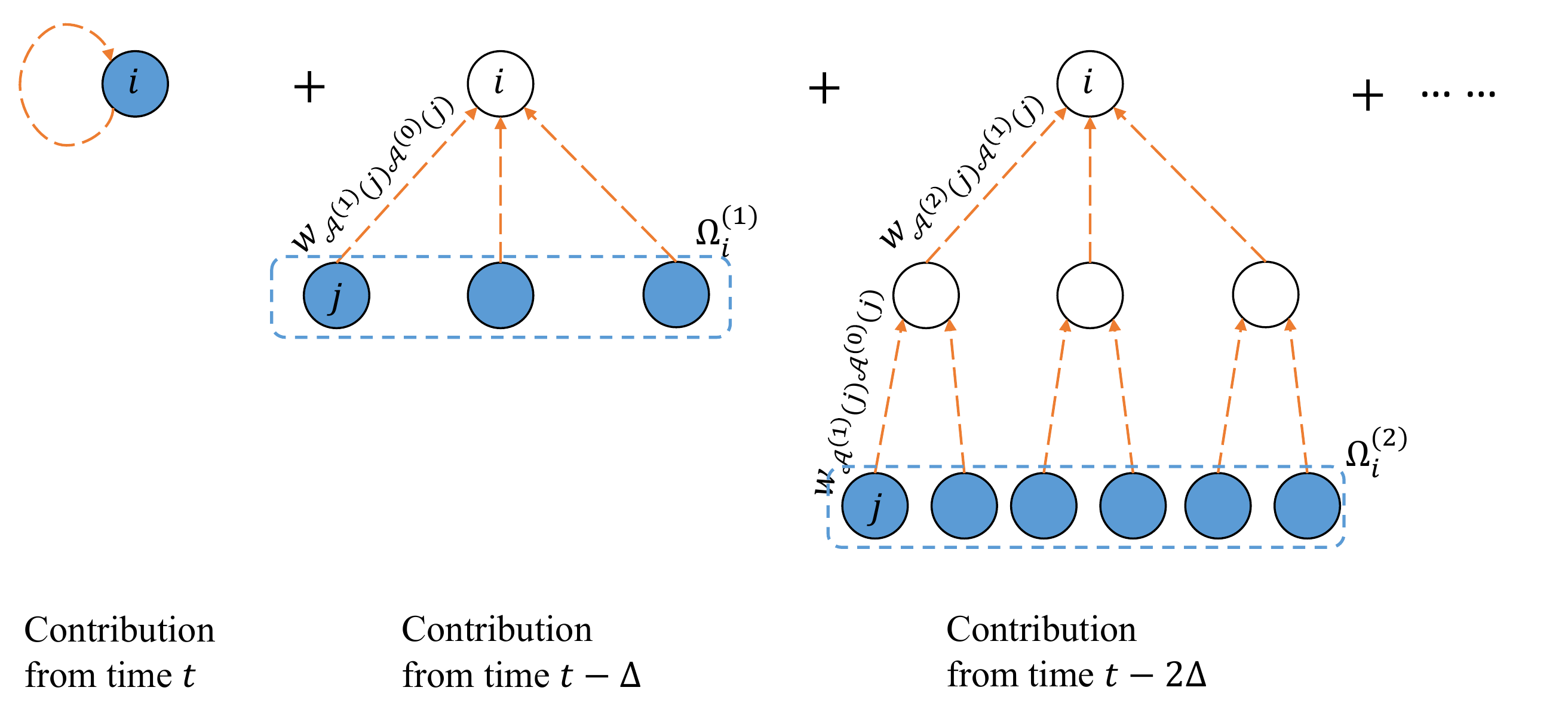}
     \vspace{-14pt}
    \caption{A graphical illustration of how the intensity $c(i,t)$ is contributed from the neighbor sets of $i$ at times $t, t-\Delta, t-2\Delta, \cdots$.}
    \label{fig:graph_new}
\end{figure}

Figure \ref{fig:graph_new} provides a graphical illustration of (\ref{eq:totalcontribution}) and the discussions above. 

Let $\log\bm{\lambda}(t) = (\log\lambda(1,t), \log\lambda(2,t), \cdots, \log\lambda(N,t))^T$ be a vector that contains the log intensity functions on all $N$ segments of a network $L$. For $k=0,1,\cdots,K$, let $\bm{c}(t-k\Delta)=(c(1,t-k\Delta), c(2,t-k\Delta), \cdots, c(N,t-k\Delta))^T$. Then, the approximated log intensity $\log\bm{\lambda}(t)$ (by retaining the first $K$ terms in the series (\ref{eq:analyticform})) admits the following matrix form:
 \vspace{-14pt}
\begin{equation}\label{eq:graphmatrix}
    \log\bm{\lambda}(t) \approx \bm{c}(t) + \xi\bm{W}\bm{c}(t-\Delta) + \cdots + \xi^K\bm{W}^K\bm{c}(t-K\Delta),
\end{equation}

 \vspace{-14pt}
\noindent where $\bm{W}$ is an $N \times N$ weight matrix with its $(i,j)$-th entry, $w_{ij}$,  being the contribution weight to $h(i,t)$ from $\log\lambda(j, t-\Delta)$, and $\bm{W}^K$ is the $K$-th power of $\bm{W}$. Note that

$\bullet$ Because the intensity on a segment $i$ is only affected by its neighboring segments, $\bm{W}$ is sparse with $w_{ij}=0$ for $j$ that is not a neighbor of segment $i$. This is similar to the idea of Convolutional Neural Network for which a node is only connected with a subset of the nodes in the previous layer. 

$\bullet$ We may understand why the matrix form (\ref{eq:graphmatrix}) holds from the perspective of the graph theory. Note that, a non-zero weight $w_{ij}^{(k)}$ in $\bm{W}^k$ implies that segment $j$ can reach segment $i$ with a $k$-step walk in the network. More importantly, the value of $w_{ij}^{(k)}$ in $\bm{W}^k$ is the sum of the contribution weights that are the multiplications of $w_{\cdot,\cdot}$ of the linked segments along all possible $k$-step walks from segment $j$ to segment $i$. Thus, the contribution of $\bm{c}(t-k\Delta)$ to $\log\bm{\lambda}(t)$ naturally admits the expression of $\xi^k\bm{W}^k\bm{c}(t-k\Delta)$. For example, $w_{ij}^{(2)} \neq 0$ in $\bm{W}^{2}$ indicates that segment $j$ can walk to segment $i$ with two steps. The value of $w_{ij}^{(2)}$ is the total contribution from all possible two-step walks from segment $j$ to segment $i$ in the network. Then, the corresponding contribution of $\bm{c}(t-2\Delta)$ to $\log\bm{\lambda}(t)$ is $\xi^2\bm{W}^k\bm{c}(t-2\Delta)$.

$\bullet$ The proposed model accepts different choices of the weight. For example, equal weight, i.e., $w_{ii'} = 1\big/|\Omega_i|$, which implies all neighboring segments of segment $i$ equally contribute to the $i$-th segment, or exponential kernel of distance, i.e., $\exp{(-d_{ii'})}\big/|\Omega_i|$, where $d_{ii'}$ is the distance between the centers of segments $i$ and $i'$. The weights may not even take any specific parametric forms but are learned from data.

\vspace{6pt}
Finally, as explained in Sec \ref{sec:model_formulation}, when $c(i,t)$ is modeled by a linear function of covariates, we obtain a linear model for $\log\bm{\lambda}(t)$ that incorporates covariate information:
 \vspace{-16pt}
\begin{equation}\label{eq:matrixintensity}
    \begin{aligned}
        \log\bm{\lambda}(t) &\approx \bm{X}(t)\bm{\beta} + \xi\bm{W}\bm{X}(t-\Delta)\bm{\beta} + \cdots + \xi^K\bm{W}^K\bm{X}(t-K\Delta)\bm{\beta}\\
                            &= \left(\sum_{k=0}^{K}\xi^k\bm{W}^k\bm{X}(t-k\Delta)\right)\bm{\beta} \triangleq \tilde{\bm{X}}(t)\bm{\beta},
    \end{aligned}
\end{equation}

 \vspace{-0pt}
\noindent where $\bm{X}(t) = (\bm{x}(1,t),\bm{x}(2,t),\cdots, \bm{x}(N,t))^T$ is the covariate matrix at time $t$, and $\tilde{\bm{X}}(t) = (\tilde{\bm{x}}(1,t),\tilde{\bm{x}}(2,t),\cdots, \tilde{\bm{x}}(N,t))^T$ is the transformed covariate matrix through the convolution operation, and we call it the Convolutional Covariate Matrix in this paper. The structure of $\bm{W}$ controls the spatio-temporal dynamics of the intensity functions over the linear network:

$\bullet$ If $\bm{W}$ is an identity matrix, the intensity function for each line segment only depends on its own historical information, and does not have spatial interaction with other line segments. 

$\bullet$ If all elements in $\bm{W}$ are zeros, there exist no historical effects nor spatial interactions for the intensity functions. In this case, the intensity functions can be completely explained by current covariates, and the proposed model (\ref{eq:intensity}) degenerates to $\log\lambda(i,t) = \bm{x}^T(i,t)\bm{\beta}$, which is widely used in existing NHPP models. 

$\bullet$ In this work, the spatio-temporal dependency is embedded in $\bm{W}$, and such dependency spans over all segments due to the $\mathcal{NC}$ operator defined in (\ref{eq:operator}). Specifically, $\{\bm{W}^m\}_{m=1}^K$ in (\ref{eq:matrixintensity}) are induced by $\mathcal{NC}$ on the linear network considering the past $K$ time steps.

Based on (\ref{eq:matrixintensity}), we can immediately obtain the log-likelihood function for the unknown parameters, including the decay factor $\xi$ and coefficients $\beta_0, \beta_1, \cdots, \beta_q$. Let $\bm{\theta}=(\xi, \beta_0, \beta_1, \cdots, \beta_q)^T$, we have \citep{peng2005space}:
  \vspace{-14pt}
\begin{equation}\label{eq:likelihood}
    \begin{aligned}
        \ell(\bm{\theta}) &= \sum_{i=1}^N \sum_{j=1}^{b_i}\log\lambda(i, t_j) - \sum_{i=1}^N\int_{0}^{T}\lambda(i, t)dt\\
                          &= \sum_{i=1}^N \sum_{j=1}^{b_i}\tilde{\bm{x}}^T(i,t_j)\bm{\beta} - \sum_{i=1}^N\int_{0}^{T}\exp\left[\tilde{\bm{x}}^T(i,t)\bm{\beta}\right]dt,
    \end{aligned}
\end{equation}

 \vspace{-14pt}
\noindent where $b_i$ is the number of events on segment $i$, and $T$ is the length of the observation period. 

\subsection{Additional Discussions: An RNN Representation}\label{sec:model_inherited_rnn}
\vspace{-8pt}
In this subsection, we present a Recurrent Neural Network representation of the proposed model. Note that, the proposed linear model in (\ref{eq:matrixintensity}) suggests that the intensity function at time $t$ depends on the historical and current covariate information. Such a structure enables us to draw a connection between the proposed model and an RNN---a feed-forward neural network with an input layer, a recurrent layer, and an output layer. 
\begin{figure}[h!]
    \centering
    \includegraphics[width=0.8\textwidth]{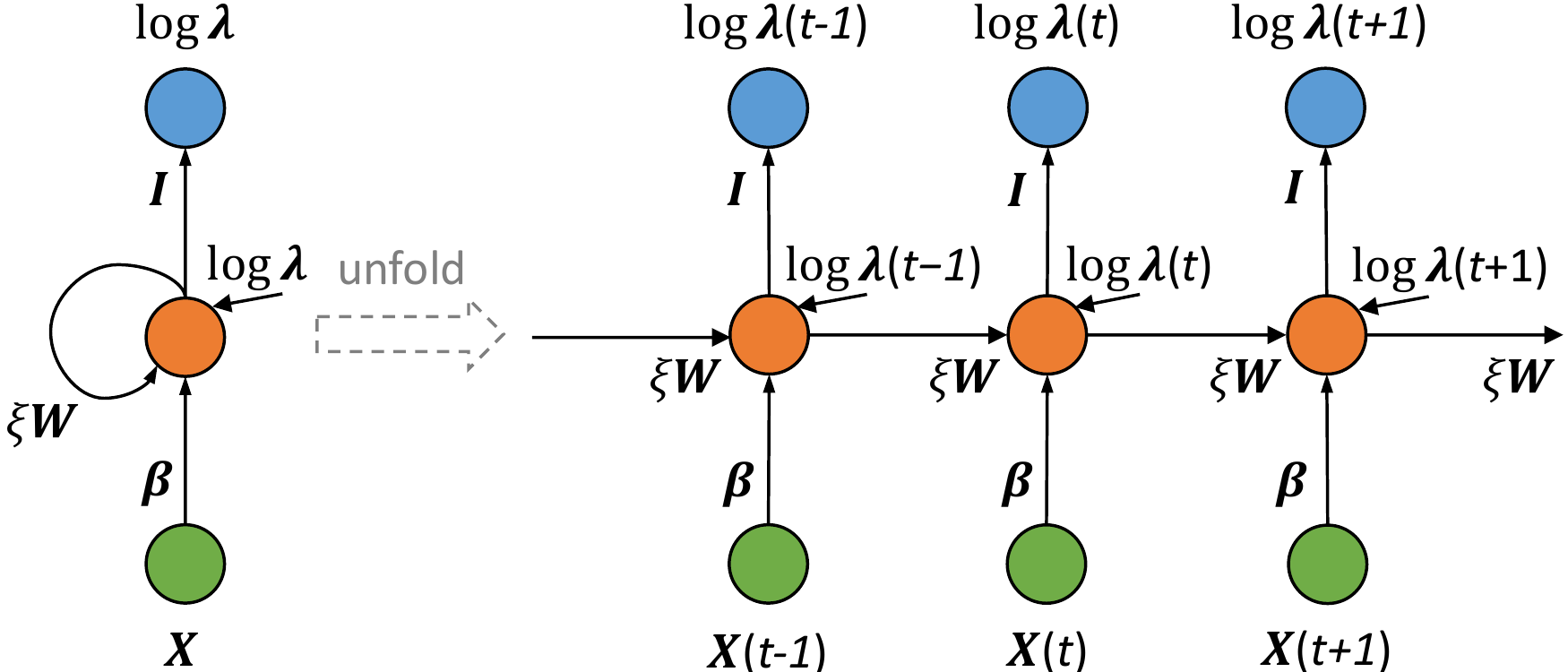}
    \caption{The RNN representation of the proposed convolutional Non-homogeneous Poisson Process model.}
    \label{fig:RNN}
\end{figure}

\vspace{-8pt}
The RNN representation of the proposed model is shown in Figure \ref{fig:RNN}. Motivated from the proposed model (\ref{eq:matrixintensity}), the input layer takes the input of the $N\times (q+1)$ covariate matrix $\bm{X}(\cdot)$, while the output layer outputs the intensity functions $\log\bm{\lambda}(\cdot)$. The recurrent layer consists of the recurrent edge and hidden state, in which the recurrent edge repeatedly feeds the previous hidden state to the current state. The unfolded version of this RNN has a many-to-many structure (i.e., many inputs and outputs) that can be represented by the following forward-propagation equations \citep{fan2021selective}:
 \vspace{-18pt}
\begin{equation}\label{eq:RNN}
        \bm{h}(t) = \xi\bm{W}\bm{h}(t-1) + \bm{X}(t)\bm{\beta}, \quad\quad
        \bm{o}(t) = \bm{I}\bm{h}(t) \triangleq \log\bm{\lambda}(t),    
\end{equation}

 \vspace{-18pt}
\noindent where $\bm{h}(t)$ and $\bm{o}(t)$ are respectively the hidden and output states at time $t$, the hidden-to-output connection $\bm{I}$ is an identity matrix in our case, and the weight vector $\bm{\beta}$ and weight matrix $\xi\bm{W}$ are the input-to-hidden and hidden-to-hidden connections respectively. It is immediately seen that, (\ref{eq:RNN}) implies that the intensity function at time $t$ is given by the sum of $\bm{X}(t)\bm{\beta}$ (i.e., the current effects of covariates) and $\xi\bm{W}\bm{h}(t-1) = \xi\bm{W}\log\bm{\lambda}(t-1)$ (i.e., the cumulative effects through spatio-temporal dependency), which is exactly the same as the proposed model (\ref{eq:matrixintensity}). Because this RNN representation is originated from the proposed model, we call it the model-inherited RNN (\texttt{mRNN}) in this paper. 

In the \texttt{mRNN} (\ref{eq:RNN}), the parameters to be learned are the same as in the model (\ref{eq:matrixintensity}), i.e., $\bm{\theta}=(\xi, \beta_0, \beta_1,\cdots,\beta_q)^T$, which can be estimated through maximizing the log-likelihood function $\ell(\bm{\theta})$ in (\ref{eq:likelihood}) using the back-propagation through time (BPTT). The BPTT provides the computational procedure for the gradients of unknown parameters. Then, the obtained gradient information can be adopted to train the RNN with the general-purpose gradient-based techniques. In particular, 

 Because the likelihood depends on $\bm{\beta}$ and $\xi$ through the hidden states $\{\bm{h}(t)\}_{t=0}^T$, we have
\begin{equation}\label{eq:grd_b}
    \frac{\partial{\ell(\bm{\theta})}}{\partial{\bm{\beta}}} = \sum_{t=0}^{T}\frac{\partial{\ell(\bm{\theta})}}{\partial{\bm{h}(t)}} \frac{\partial{\bm{h}(t)}}{\partial{\bm{\beta}}} = \sum_{t=0}^{T}\bm{X}^T(t)\frac{\partial{\ell(\bm{\theta})}}{\partial{\bm{h}(t)}}
\end{equation}
\begin{equation}\label{eq:grd_xi}
    \frac{\partial{\ell(\bm{\theta})}}{\partial{\xi}} = \sum_{t=0}^{T}\frac{\partial{\ell(\bm{\theta})}}{\partial{\bm{h}(t)}} \frac{\partial{\bm{h}(t)}}{\partial{\xi}} = \sum_{t=0}^{T}\bm{h}^T(t-1)\bm{W}^T\frac{\partial{\ell(\bm{\theta})}}{\partial{\bm{h}(t)}}.
\end{equation}

To compute $\{\frac{\partial{\ell(\bm{\theta})}}{\partial{\bm{h}(t)}}\}_{t=0}^{T}$ in (\ref{eq:grd_b}) and (\ref{eq:grd_xi}), the following back-propagation can be adopted through time: 

$\bullet$ At the final time step $T$, the likelihood function depends on the hidden state $\bm{h}(T)$ only via $\bm{o}(T)$, then we have
        \begin{equation}\label{eq:recurrence1}
            \frac{\partial{\ell(\bm{\theta})}}{\partial{\bm{h}(T)}} =  \frac{\partial{\ell(\bm{\theta})}}{\partial{\bm{o}(T)}}\frac{\partial{\bm{o}(T)}}{\partial{\bm{h}(T)}} = \frac{\partial{\ell(\bm{\theta})}}{\partial{\bm{o}(T)}}.
        \end{equation}
        
$\bullet$  At the previous time step $t$ before $T$, the likelihood function depends on the hidden state $\bm{h}(t)$ through $\bm{o}(t)$ and $\bm{h}(t+1)$, then we have
 \vspace{-10pt}
        \begin{equation}\label{eq:recurrence2}
            \begin{aligned}
                \frac{\partial{\ell(\bm{\theta})}}{\partial{\bm{h}(t)}} &= \frac{\partial{\ell(\bm{\theta})}}{\partial{\bm{o}(t)}}\frac{\partial{\bm{o}(t)}}{\partial{\bm{h}(t)}} + \frac{\partial{\ell(\bm{\theta})}}{\partial{\bm{h}(t+1)}}\frac{\partial{\bm{h}(t+1)}}{\partial{\bm{h}(t)}}\\
                &= \frac{\partial{\ell(\bm{\theta})}}{\partial{\bm{o}(t)}} + \xi\bm{W}^T\frac{\partial{\ell(\bm{\theta})}}{\partial{\bm{h}(t+1)}}.
            \end{aligned}
        \end{equation}
  
   \vspace{-8pt}
$\bullet$  Using the recurrence relation in (\ref{eq:recurrence2}) and (\ref{eq:recurrence1}), $\{\frac{\partial{\ell(\bm{\theta})}}{\partial{\bm{h}(t)}}\}_{t=0}^{T}$ can be computed by 
 \vspace{-14pt}
        \begin{equation}
            \frac{\partial{\ell(\bm{\theta})}}{\partial{\bm{h}(T-n)}}=\sum_{i=0}^n(\xi\bm{W}^T)^i\frac{\partial{\ell(\bm{\theta})}}{\partial{\bm{o}(T-n+i)}}.
        \end{equation}

Compared with the proposed model, the RNN representation presents two main advantages and one potential limitation:

(Advantages) The first advantage is that the structure of the \texttt{mRNN} in Figure \ref{fig:RNN} is more flexible and can be adapted or extended as needed. For example, $\bm{W}$ is a sparse weighted adjacency matrix that determines the spatial dependency. In the proposed statistical model, certain parametric assumptions on the structure of $\bm{W}$ are often imposed. However, the spatial dependency (i.e., the weights associated with the hidden-to-hidden connection) can be directly learned from the \texttt{mRNN} without imposing parametric assumptions on  $\bm{W}$. 

The second advantage of the RNN representation is that the computational cost can potentially be lower when the \texttt{mRNN} is adopted. Based on the structure shown in Figure \ref{fig:RNN}, we see that the output of $\{\log\bm{\lambda}(t)\}_{t=0}^{T}$ requires the multiplications of $\xi\bm{W}$ with $\log\bm{\lambda}(t)$ as well as $\bm{X}(t)$ with $\bm{\beta}$ in the \texttt{mRNN}. In proposed statistical model, on the other hand, because $\log\bm{\lambda}(t)=\tilde{\bm{X}}(t)\bm{\beta}$ in (\ref{eq:likelihood}) is used to compute $\{\log\bm{\lambda}(t)\}_{t=0}^{T}$, the computation of the convolutional matrix $\tilde{\bm{X}}$ requires the multiplication of $\xi^k\bm{W}^k$ with $\bm{X}(t-k\Delta)$. This is computational more expensive than the multiplication between the matrices (i.e., $\xi\bm{W}$ and $\bm{X}(t)$) and the vectors (i.e., $\log\bm{\lambda}(t)$ and $\bm{\beta}$) in the  \texttt{mRNN}.

\begin{figure}[h!]
    \centering
    \includegraphics[width=0.4\textwidth]{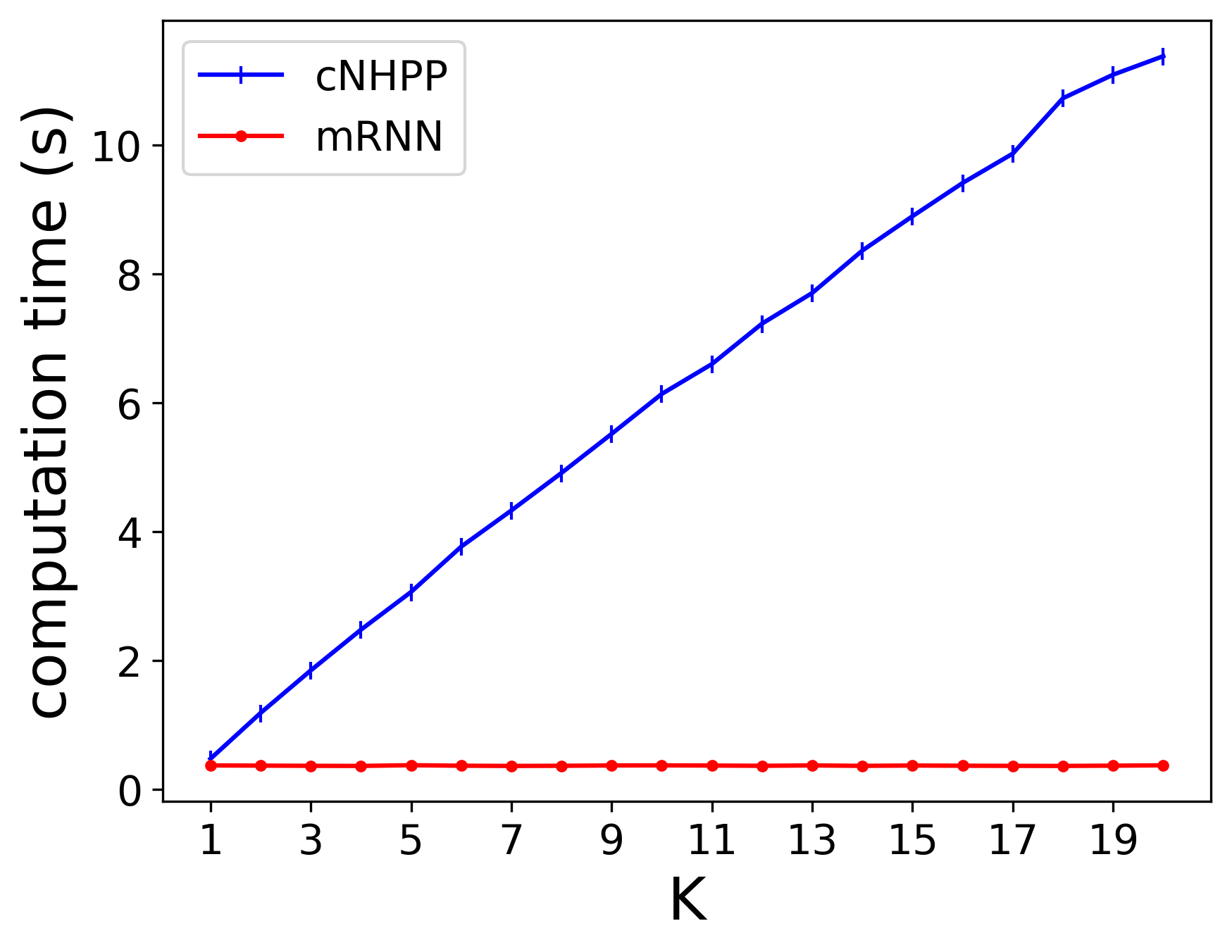}
     \vspace{-10pt}
    \caption{Comparison of the computational time of evaluating $\{\log\bm{\lambda}(t)\}_{t=0}^{T}$ respectively based on the proposed statistical model and its RNN representation.}\label{fig:ct_cp}
\end{figure}

For illustrative purposes, we compare the computational time of evaluating $\{\log\bm{\lambda}(t)\}_{t=0}^{T}$ respectively based on the proposed statistical model and its RNN representation. The \texttt{mRNN} is implemented in \texttt{PyTorch} and the multiplications are performed with the tensor data type. We also compute $\{\log\bm{\lambda}(t)\}_{t=0}^{T}$ with the tensor data type for the proposed approach.
The dimensions of $\bm{W}$ and $\bm{X}(t)$ are kept the same as in Section \ref{sec:results}, and the comparison result is shown in Figure \ref{fig:ct_cp}. We see that the computational time for the proposed model increases almost linearly with the increase of the truncation number $K$. Even for $K=1$, the computation cost of the proposed model is still higher than that of the \texttt{mRNN}. In the application example presented in Section \ref{sec:results}, we set $K=7$ and the computational time associated with the statistical model is approximately 12 times longer than using its RNN representation. This strongly demonstrates the potential of the \texttt{mRNN} in directly learning the spatial dependency of $\bm{W}$, provided that there are sufficient data for training the model.  

\vspace{-8pt}
(Limitation) Finally, it is important to mention that training the \texttt{mRNN} requires a large event data set. In the wildfire application considered in this paper, wildfire are still considered as rare events in the sense that majority of the transmission lines do not experience any fire events. As shown in the next section, 15 fire events are reported among 6398 power transmission lines within a one-month window. Although 15 events are already considered significant from the power grid operation and maintenance perspective, it is still a very small number for  training the \texttt{mRNN} . For this reason, the \texttt{mRNN} becomes less effective in learning the spatial dependence structure among different transmission lines in the example presented in the next section. In this case, the proposed statistical model (\ref{eq:matrixintensity}), with some pre-specified assumption on the spatial dependence structure, is found to provide better results.

\section{Application: Wildfire Risks on Networks of Power Transmission Lines in California} \label{sec:application}
We apply the proposed approach for modeling and predicting wildfire risks on networks of power transmission lines in parts of California. Section \ref{sec:data_description} provides detailed descriptions of the data obtained for this application. In Section \ref{sec:results}, we present and discuss the model outputs and provide some useful insights on the impact of wildfires on power delivery infrastructures. Comparison studies are performed in Section \ref{sec:comparison}. 

\vspace{-8pt}
\subsection{Data}\label{sec:data_description}
\vspace{-8pt}
This application example involves four major datasets: (i) power delivery infrastructure data, (ii) wildfire incident data, (iii) meteorological data, and (iv) vegetation data. 

\begin{figure}[H]
    \centering
    \includegraphics[width=1\textwidth]{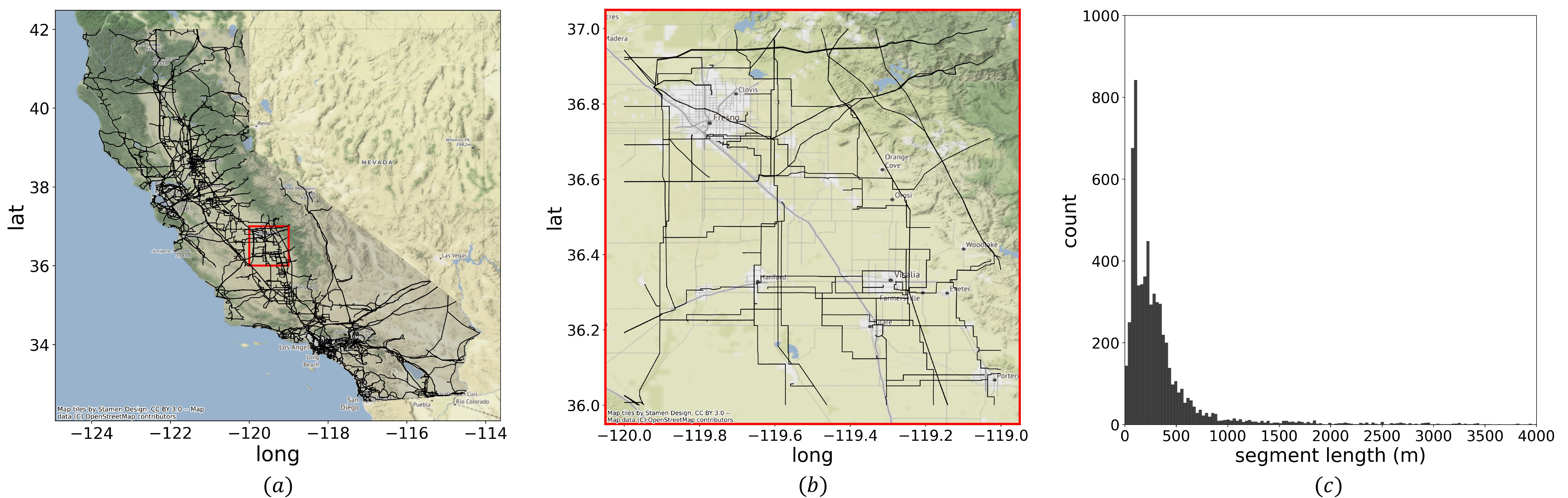}
    \caption{(a) Main power transmission lines in California; (b) Power transmission lines in the study area; (c) Histogram of the  length of line segments in the study area.}
    \label{fig:TransmissionLines}
\end{figure}

Figure \ref{fig:TransmissionLines}(a) shows the main power transmission lines in California. This dataset is obtained, in the shapefile format, from the U.S. Energy Information Administration (EIA). In particular, we focus on a spatial area indicated by the square in Figure \ref{fig:TransmissionLines}(a). This spatial area is defined by $[120^{\circ}\text{W}, 119^{\circ}\text{W}]\times[36^{\circ}\text{N}, 37^{\circ}\text{N}]$, and there is a total number of 6398 power transmission lines within this area; see Figure \ref{fig:TransmissionLines}(b). Figure \ref{fig:TransmissionLines}(c) shows the histogram of the lengths of these 6398 segments, and most of these line segments are less than 1000 meters. 

\vspace{-16pt}
The fire incident dataset contains the information about overhead power-line fires, including fire locations, dates and corresponding power-line contents. The wildfire incident data are obtained from the California Public Utilities Commission (CPUC), which is a government agency that regulates public utility companies including privately owned electric, natural gas, and telecommunications companies. In response to increasingly severer overhead power-line wildfires, CPUC requires electricity companies to report data on power-line fires. We obtain the fire incident data reported from the Southern California Edison (SCE), Pacific Gas and Electric (PG\&E), and San Diego Gas and Electric (SDG\&E) companies. From Jun 1 to Jun 30, 2019, a total number of 15 wildfires were reported within the study area. On average, there was a fire incident every two days due to the high temperature and dry weather at the beginning of the summer season. 
\begin{figure}[H]
    \centering
    \includegraphics[width=1\textwidth]{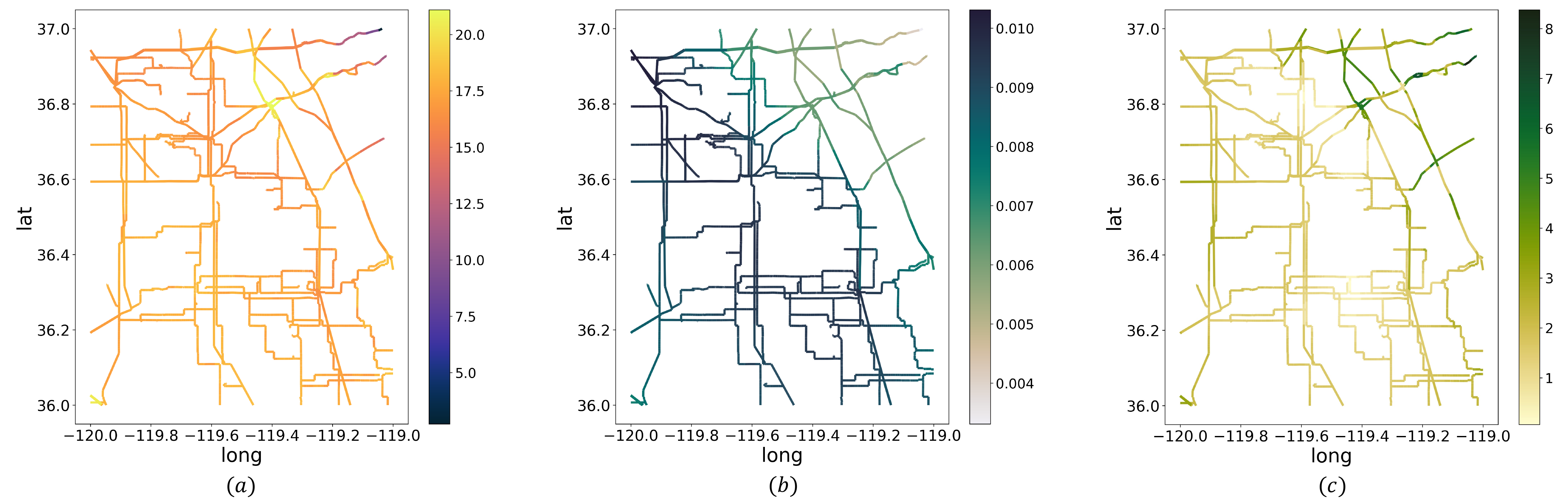}
    \caption{Illustration of three meteorological variables from the HRRR model on Jun 01, 2019: (a) TMP ($^{\circ}$C); (b) SPFH ($\text{kg}\cdot\text{kg}^{-1}$); (c) WIND ($\text{m}\cdot\text{s}^{-1}$).}
    \label{fig:noaa}
\end{figure}

\vspace{-16pt}
Meteorological data are obtained from the High-Resolution Rapid Refresh (HRRR) model maintained by the National Oceanic and Atmospheric Administration (NOAA). The HRRR model provides hourly meteorological data with a spatial resolution of 3 kilometers. Although HRRR contains 170 meteorological variables in 2D surface levels, most of these variables are the same meteorological conditions but in different pressure regions. Hence, we select three representative variables, including temperature (2 meters above ground), specific humidity (2 meters above ground), and wind speed (10 meters above ground). For convenience, we denote temperature as TMP ($^{\circ}$C), specific humidity as SPFH ($\text{kg}\cdot\text{kg}^{-1}$), and wind speed as WIND ($\text{m}\cdot\text{s}^{-1}$). Because power lines do not always locate in a regular grid, the meteorological data at the nearest grid points are assigned to each power line segment. As an illustration, Figure \ref{fig:noaa} shows the meteorological conditions on Jun 01, 2019.

Vegetation data are obtained from the Normalized Difference Vegetation Index (NDVI) that reflects the vegetation-water status. A higher NDVI corresponds to a denser and healthier vegetation canopy and vice versa. 
This dataset is obtained from the Moderate Resolution Imaging Spectroradiometer (MODIS) on board NASA's Aqua and Terra satellites. Because MODIS only has 8-day NDVI data products, we manually calculate daily NDVI using the daily land surface reflectance products using the following equation: 
\vspace{-8pt}
\begin{equation}\label{eq:ndvi}
    \text{NDVI} = \frac{\rho_\text{NIR}-\rho_\text{red}}{\rho_\text{NIR}+\rho_\text{red}},
\end{equation}

\begin{figure}[h!]
    \centering
    \includegraphics[width=0.8\textwidth]{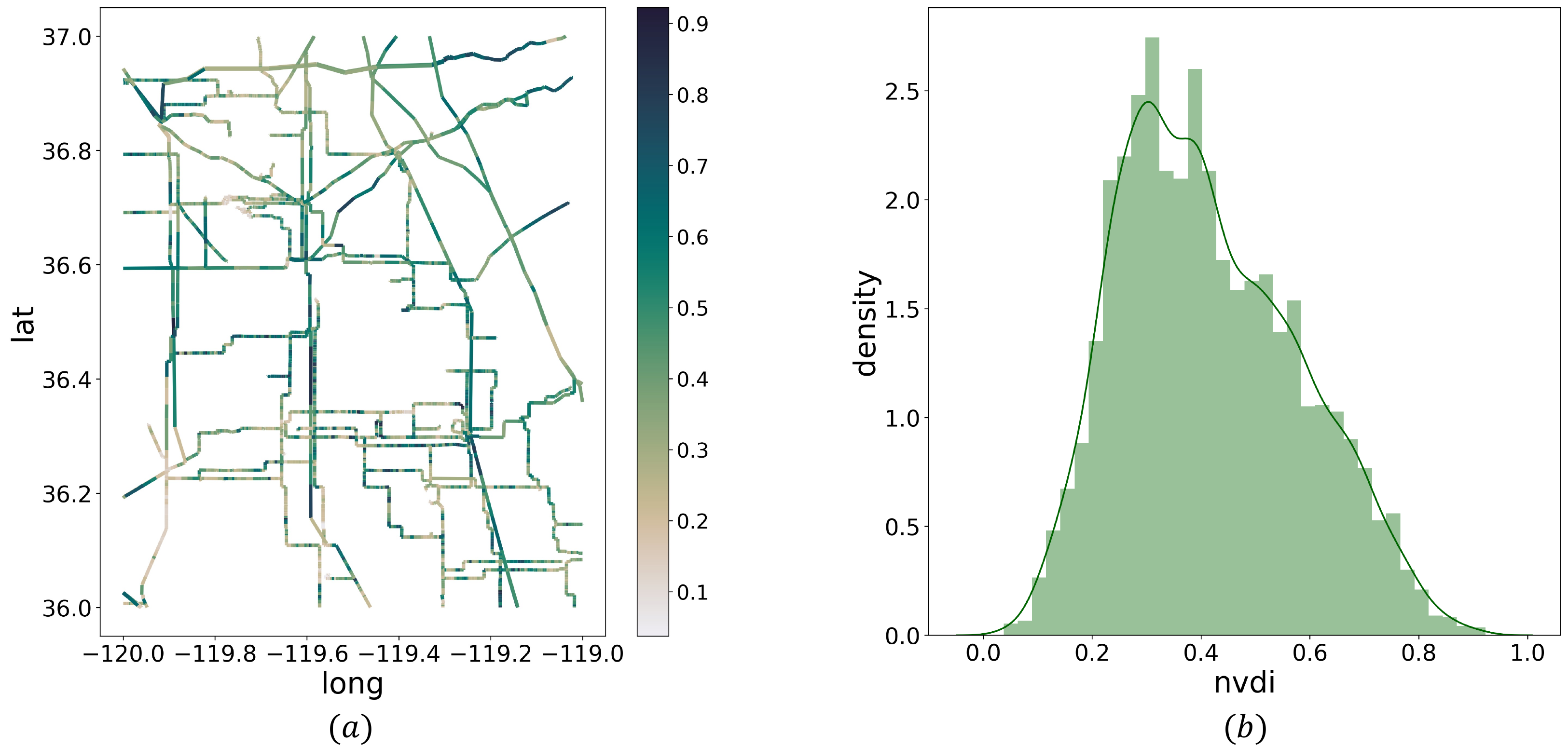}
    \caption{Illustration of the processed NDVI data from MODIS on Jun 1, 2019: (a) NDVI data projected on power transmission lines; (b) NDVI density plot.}
    \label{fig:nasa}
\end{figure}

\vspace{-8pt}
\noindent where $\rho_\text{red}$ and $\rho_\text{NIR}$ respectively denote the reflectances of near-infrared and red spectral regions. Detailed descriptions about the reflectance products can be found in \citet{MODIS}. Here, $\rho_\text{red}$ and $\rho_\text{NIR}$ have a 250-m spatial resolution in MODIS land surface reflectance products, and so do the processed NDVI data. 
Figure \ref{fig:nasa} gives an example of the processed NDVI values that range from 0 to 1. 

\vspace{-8pt}
\subsection{Results and Discussions}\label{sec:results}
\vspace{-8pt}
Based on the power transmission line data above, $\bm{\lambda}(t)$ is a column vector that contains the intensity functions of the 6398 power lines. The convolutional covariate matrix $\tilde{\bm{X}}$ is a $5\times6398$ matrix, and the sparse weighted adjacency matrix $\bm{W}$ has a dimension of $6398\times6398$. In this application, we let $w_{ii'} = 1\big/|\Omega_i|$ for $i'\in\Omega_i$, implying that all neighbors of segment $i$ equally contribute to the fire intensity of the $i$-th segment. As discussed in Section \ref{sec:model_inherited_rnn}, such a parametric assumptions on the structure of $\bm{W}$ may not be needed when the \texttt{mRNN} is used, which has the capability of directly learning the entries of $\bm{W}$ (i.e., the weights associated with the hidden-to-hidden connection) from the training data. 

We let $x_1(i,t)$, $x_2(i,t)$, $x_3(i,t)$ and  $x_4(i,t)$ respectively denote the NDVI, TMP, WIND, and SPFH for the $i$-th segment at day $t$. All covariates are standardized so as to facilitate the comparison between their effects $\beta_1$, $\beta_2$, $\beta_3$, and $\beta_4$. An intercept $\beta_0$ is also included. 
The parameters $\bm{\theta}=(\xi, \beta_0, \beta_1, \beta_2, \beta_3, \beta_4)^T$ in (\ref{eq:matrixintensity}) are estimated by maximizing the log-likelihood function (\ref{eq:likelihood}). Note that, the log-likelihood function (\ref{eq:likelihood}) is concave with respect to $\bm{\beta}$, but the computational cost of $\tilde{\bm{X}}(t)$ for different values of $\xi$ is extremely high. Hence, simultaneously estimating all parameters in $\bm{\theta}$ is an inefficient process. Here, we adopt a practical solution by considering a finite number of $\xi$ (for computing $\tilde{\bm{X}}(t)$), and compare the corresponding maximized log-likelihoods with respect to $\hat{\bm{\beta}}$. In the computation of $\tilde{\bm{X}}(t)$, we let the truncation number $K=7$. 

\begin{figure}[h!]
    \centering
    \includegraphics[width=0.5\textwidth]{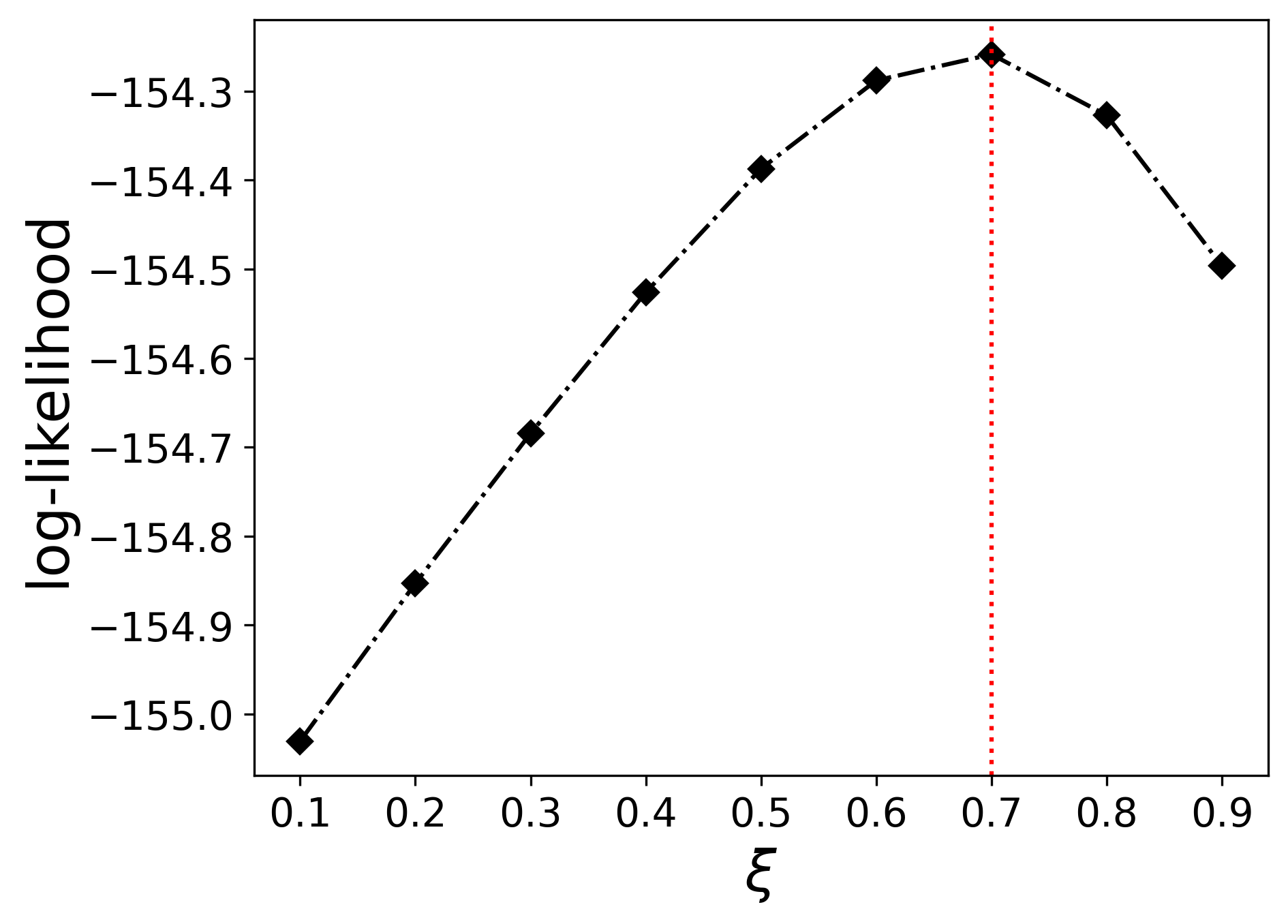}
    \vspace{-8pt}
    \caption{The maximized log-likelihood of $\bm{\beta}$ for given values of $\xi$.}
    \label{fig:md_est_plot}
\end{figure}

Figure \ref{fig:md_est_plot} shows the maximized log-likelihood for different values of $\xi$ ranging from 0 to 1. It is seen that the maximum log-likelihood is attained when $\xi=0.7$,  where $\hat{\bm{\beta}}$ is obtained by the \texttt{SciPy} package using the \texttt{L-BFGS-B} method in Python. Hence, we fix  $\hat{\xi}=0.7$, and the ML estimates $\hat{\bm{\beta}}$ are: (Intercept) $\hat{\beta}_0=-2.748$, (NDVI)  $\hat{\beta}_1= -1.226$, (TMP) $\hat{\beta}_2 = 0.661$, (WIND) $\hat{\beta}_3=0.887$, (SPFH) $\hat{\beta}_4= -0.664$. 
It is seen that higher temperature and stronger wind speed increase the wildfire risks by the factors of 1.94 (i.e., $\exp(\hat{\beta}_2)$) and 2.43 (i.e., $\exp(\hat{\beta}_3)$). Such findings can be well justified as follows: (i) a higher temperature makes the ignition of the underlying fuels more easily (e.g. grasses, shrubs, dead leaves, etc.), (ii) with an increased wind speed, the electrical conductors and surrounding vegetation are more likely to result in arcing, increasing the probability of wildfire ignition \citep{mitchell2013power, vazquez2022wildfire}. If wind speed  suddenly surges, more attention should be given to power lines under strong wind conditions, e.g., gale.


\vspace{-8pt}
Both the NDVI and the SPFH have negative impacts on wildfire risks, while SPFH has a relatively weaker effect compared with that of NDVI. Note that, a higher NDVI indicates healthier vegetation with more water conditions and fewer potential fuels that can be ignited. Similarly, a high SPFH attaches potential fuels with more moisture, reducing the wildfire risk. In fact, NDVI is found to be the most influential factor with $\hat{\beta}_1=-1.226$ (recall that all covariates are standardized). This implies that the wildfire risk is largely determined by the underlying NDVI, which is included as one of the physics-based indexes by the Keetch–Byram Drought Index (KBDI), Fire Potential Index (FPI), and Normalized Difference Water Index (NDWI) \citep{verbesselt2006evaluating, huesca2014modeling, brown2021us}. 
Because the NDVI takes the lead in these factors affecting wildfire risks, activities related to weeding are highly recommended to eliminate unhealthy vegetation around power delivery infrastructures. 
\begin{figure}[h!]
    \includegraphics[width=1\textwidth]{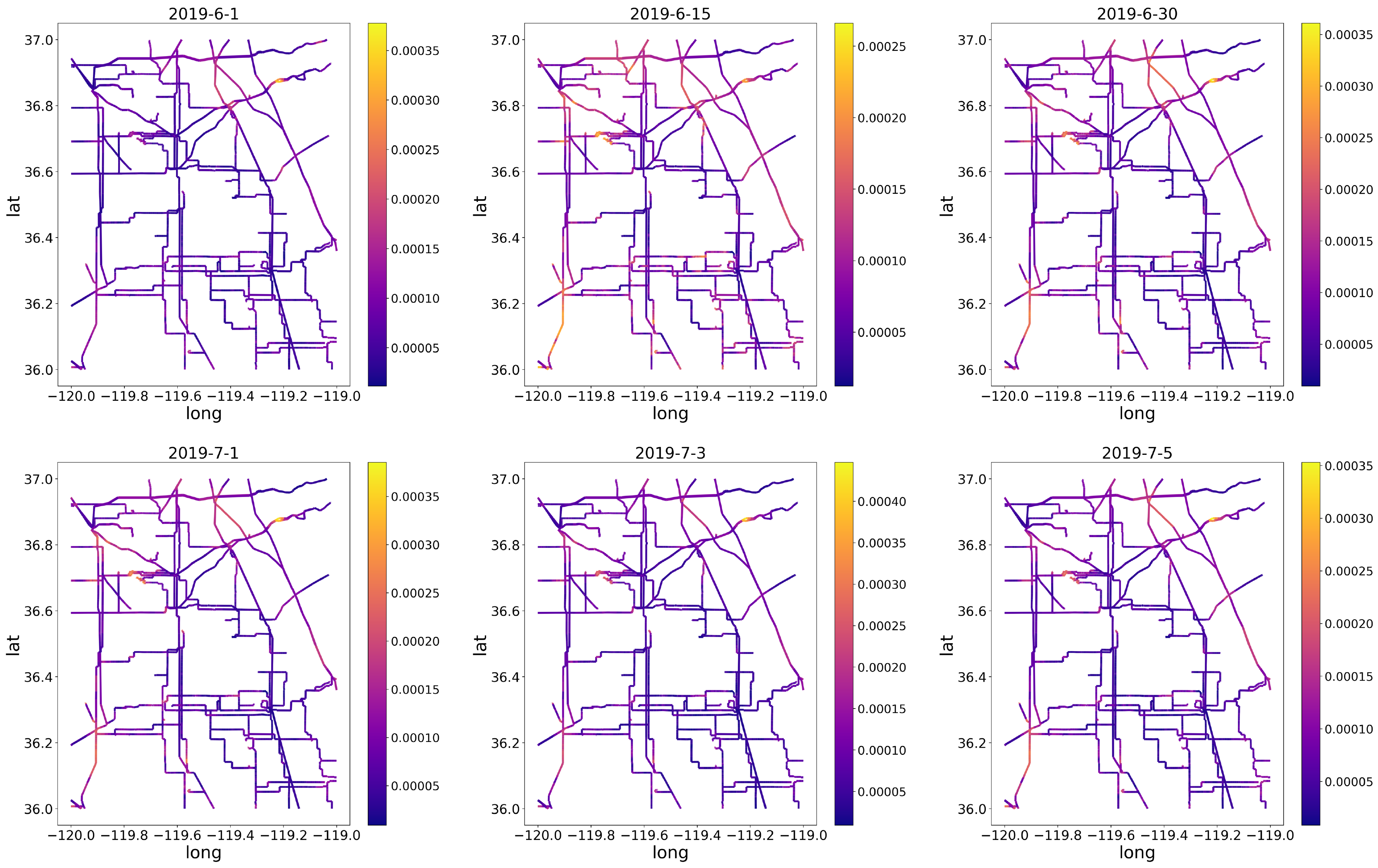}
    \caption{Estimated (top row) and predicted (bottom row) wildfire event intensities for power lines at selected days based on our proposed approach.}\label{fig:est_pred_risk}
\end{figure}

The proposed model can also be used for short-term predictions of wildfire intensities. The predicted wildfire intensities are obtained once the convolutional matrix $\tilde{\bm{X}}(t+k)$ can be computed given future covariates values, i.e., $\log\hat{\bm{\lambda}}(t+k)=\tilde{\bm{X}}(t+k)\hat{\bm{\beta}}$. Figure \ref{fig:est_pred_risk} shows the estimated and predicted wildfire event intensities based on the proposed approach. It is seen that different power line segments are associated with different wildfire risks  due to the spatially- and temporally-varying covariate information. It is also seen from the second row of Figure \ref{fig:est_pred_risk} that the predicted wildfire risks change smoothly over time. This is because the proposed model incorporates the cumulative long-term effects of covariates and the estimated decay factor $\hat{\xi}=0.7$. As a result, the wildfire intensities do not dramatically change in a short period even if the current covariates change abruptly. Note that, if a smaller decay value $\xi$  is obtained, the cumulative effects decay faster and the predicted wildfire risks are more sensitive to current environmental conditions.

\vspace{-8pt}
\subsection{Comparison and Discussions}\label{sec:comparison}
\vspace{-8pt}
We further compare the proposed \texttt{cNHPP} model with the following three models:

$\bullet$ \texttt{HPP}: Homogeneous Poisson Process (HPP) that models the wildfire intensity as a constant over the entire network, i.e., $\log\lambda(i,t) = \log\lambda$ for $i=1,2,\cdots,N$. In this case, the ML estimate of the intensity $\hat{\lambda}$ has a closed-form expression. 

$\bullet$ \texttt{NHPP}: Conventional NHPP model for which the wildfire intensity is only determined by the current covariates without accounting for the cumulative effects of covariates and spatial dependency \citep{trilles2013integration}, i.e., $\log\lambda(i,t)=\bm{x}^T(i,t)\bm{\beta}$ for $i=1,2,\cdots,N$.  In this case, the log-likelihood is concave with respect to the unknown parameters $\bm{\beta}$, making the search for the ML estimates easier than the proposed \texttt{cNHPP}. As mentioned earlier, when all elements in the weighted adjacency matrix $\bm{M}$ are zeros as in \texttt{NHPP}, our proposed \texttt{cNHPP}  degenerates to an \texttt{NHPP} model.   

$\bullet$ \texttt{mRNN}: The model-inherited RNN  described in Section \ref{sec:model_inherited_rnn}.  We implement the RNN with \texttt{PyTorch}, and employ the Adam optimizer (learning rate = 0.001) to train the model. Convergence of the loss function and unknown parameters are provided in the Supplementary Material, where a total number of $20,000$ epochs are trained.

Table \ref{tab:model_comparison} presents the estimated model parameters using the same training data described in Section \ref{sec:results}. Note that, the \texttt{HPP} model does not incorporate the covariate information that accounts for the heterogeneity of wildfire risks. 
We see from  Table \ref{tab:model_comparison} that \texttt{cNHPP} and \texttt{mRNN} have larger log-likelihood than that of \texttt{HPP} and \texttt{NHPP}. Although \texttt{NHPP},  \texttt{cNHPP} and \texttt{mRNN} yield different estimated values for the effects $\bm{\beta}$, the signs of these estimates remain consistent. This implies that considering the cumulative effects in the proposed model does not change the underlying correlation structure nor the interpretations of the relationship between the covariates and wildfire risks. It is also seen that the decay factor $\xi$ estimated by \texttt{mRNN} is smaller than that in \texttt{cNHPP}. This is because the \texttt{mRNN} naturally incorporates all historical covariate information, while \texttt{cNHPP} only considers the covariate information in the past week due to truncation ($K=7$ in our numerical example). As a result, \texttt{mRNN} obtains a smaller $\xi$ that makes the cumulative effects decay faster. 
\begin{table}[h!]
    \centering
    \caption{Estimated parameters from different models.}\label{tab:model_comparison}
    \vspace{-8pt}
    \begin{tabular}{c|c|c|c|c} 
        \Xhline{1pt}
                    & \texttt{HPP} & \texttt{NHPP} & \texttt{mRNN} & \texttt{cNHPP}       \\
        \Xhline{1pt}
        $\hat{\lambda}$ (rate$\times10^{-5}$ )   &7.815   & -              &-              & -      \\
        $\hat{\xi}$ (decay)          & -            & -              & 0.678         & 0.7        \\
        $\hat{\beta_0}$ (intercept)  & -            & -8.863         & -2.823        & -2.748       \\
        $\hat{\beta_1}$ (NDVI)       & -            & -2.723         & -1.208        & -1.226       \\
        $\hat{\beta_2}$ (TMP)        & -            & 0.704          & 0.962         & 0.661        \\
        $\hat{\beta_3}$ (WIND)       & -            & 1.344          & 0.738         & 0.887        \\
        $\hat{\beta_4}$ (SPFH)       & -            & -0.511         & -0.861        & -0.664       \\
        log-likelihood               & -156.853     & -155.217       & -154.236      & -154.259     \\
        \Xhline{1pt}
\end{tabular} 
\end{table}

\vspace{-8pt}
\begin{figure}[h!]
	\begin{subfigure}[t]{0.31\linewidth}
	\centering
	\includegraphics[width=1\textwidth]{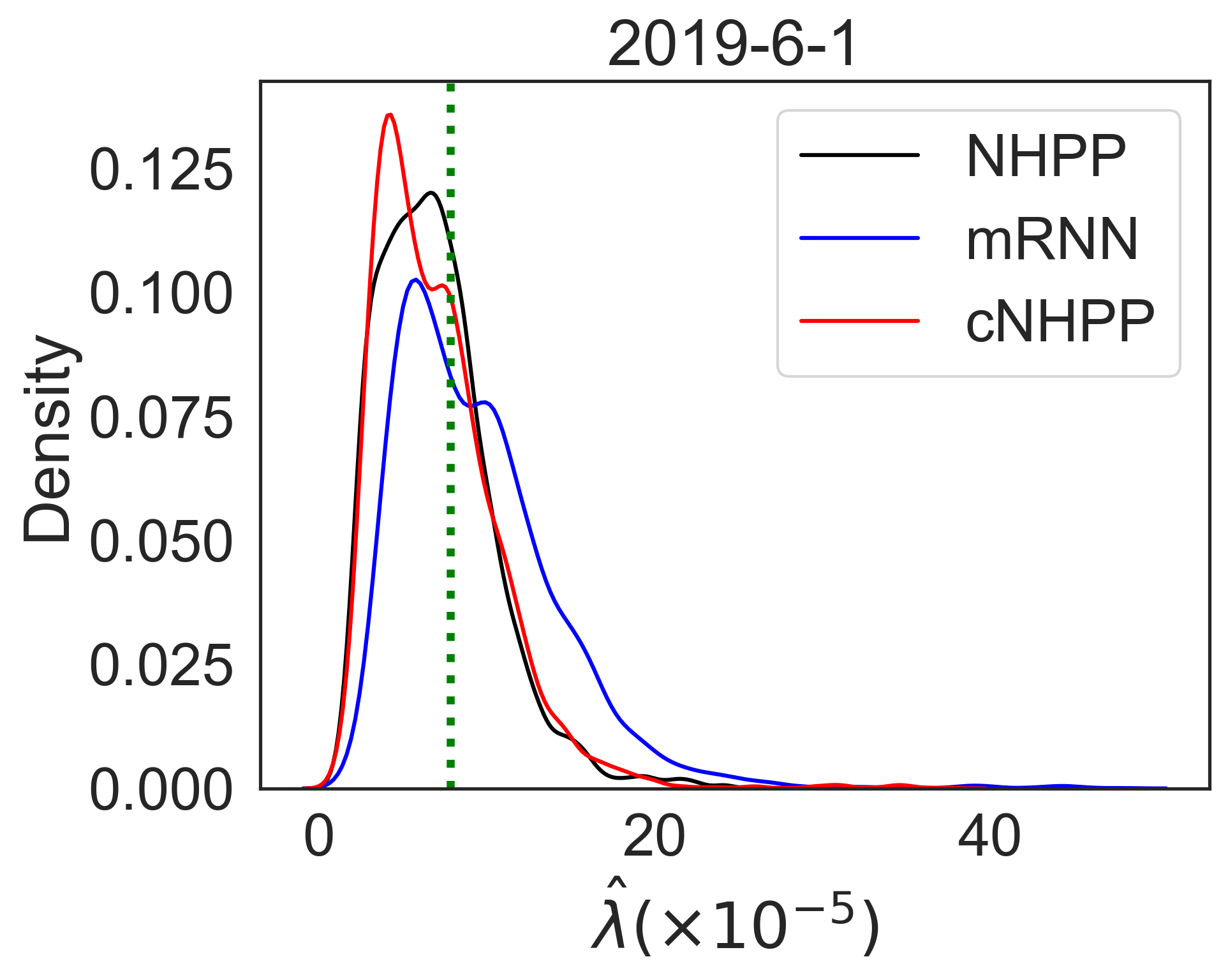}
	\end{subfigure}
	\begin{subfigure}[t]{0.31\linewidth}
    \includegraphics[width=1\textwidth]{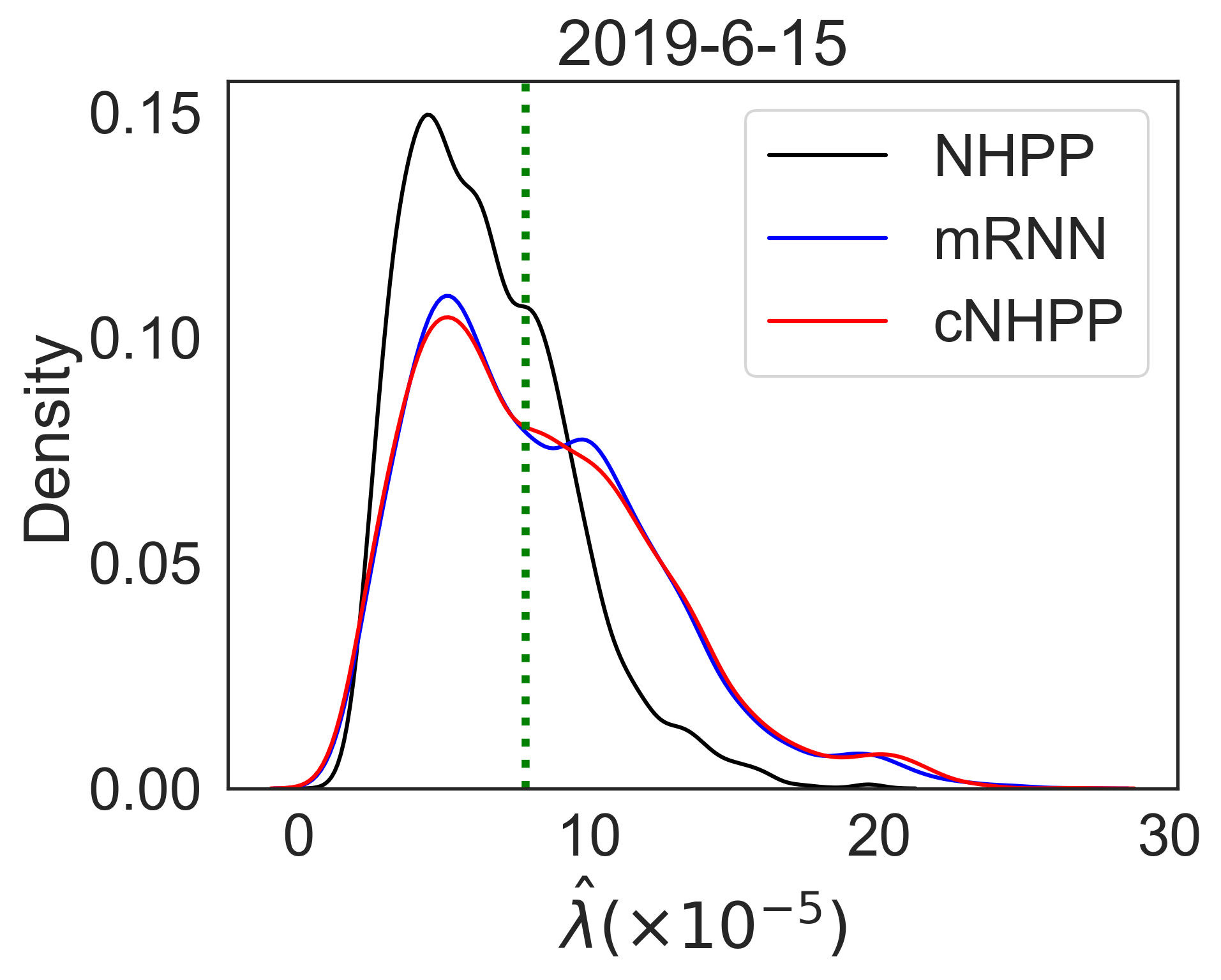}
	\end{subfigure}
	\begin{subfigure}[t]{0.31\linewidth}
	\centering
	\includegraphics[width=1\textwidth]{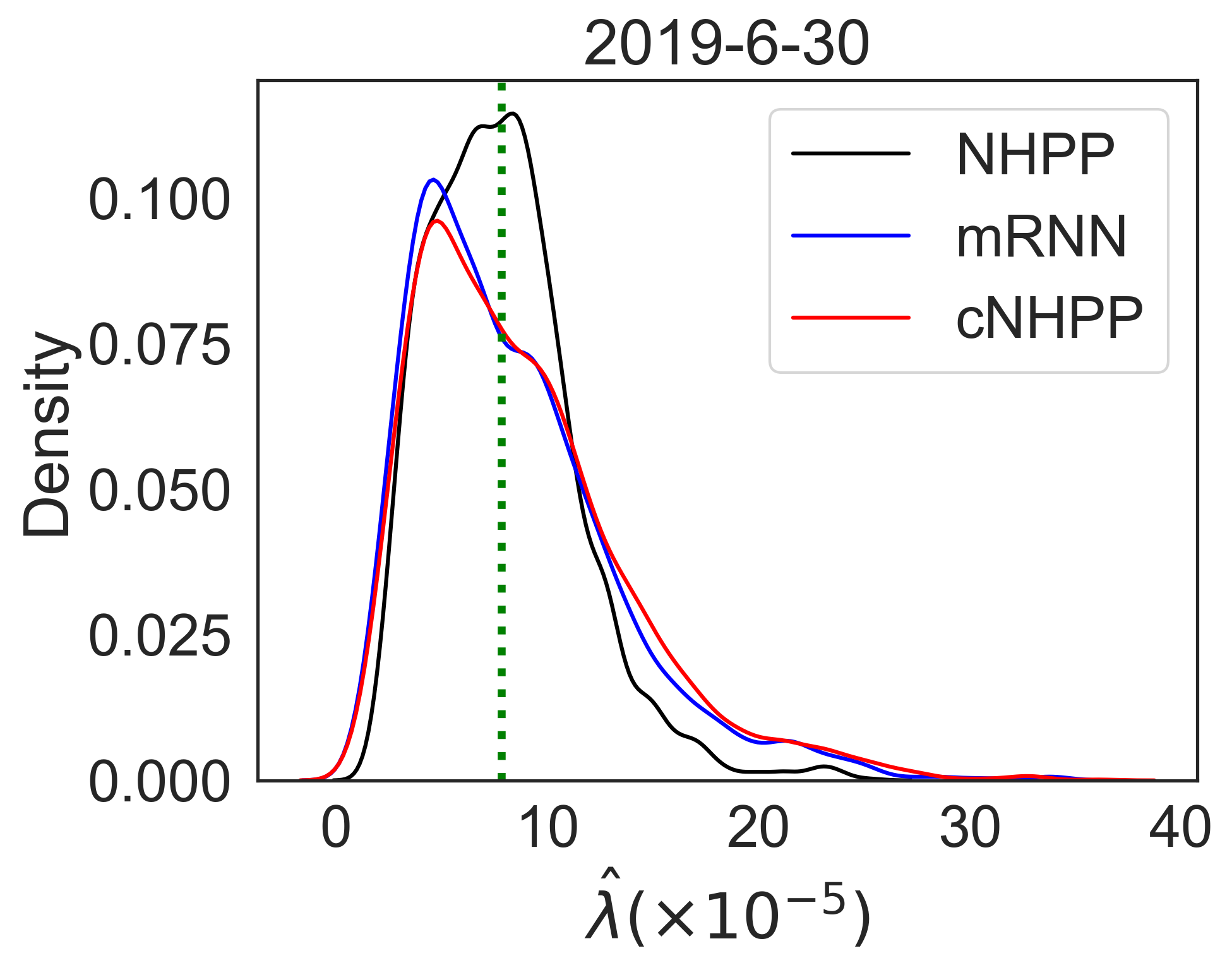}
	\end{subfigure}
	\caption{Distribution of estimated wildfire intensities on power transmission lines by different models at selected days (the dashed line denotes the estimated average wildfire intensity based on the \texttt{HPP} model).}\label{fig:est_cp}
\end{figure}

\vspace{-8pt}
Based on the estimates in Table \ref{tab:model_comparison}, we obtain the estimated intensity functions for all power-line segments from \texttt{NHPP}, \texttt{mRNN} and \texttt{cNHPP}. Figure \ref{fig:est_cp}  shows the corresponding density plots of the estimated wildfire intensities over the power lines on selected days. It is seen that density plots are all centered around the average $\hat{\lambda}$ obtained from the \texttt{HPP} model (the vertical dashed line), suggesting that all three approaches perform reasonably well in terms of estimating the underlying wildfire risks over the network of power transmission lines.

\vspace{-2pt}
\begin{figure}[h!]
    \centering
    \includegraphics[width=0.75\textwidth]{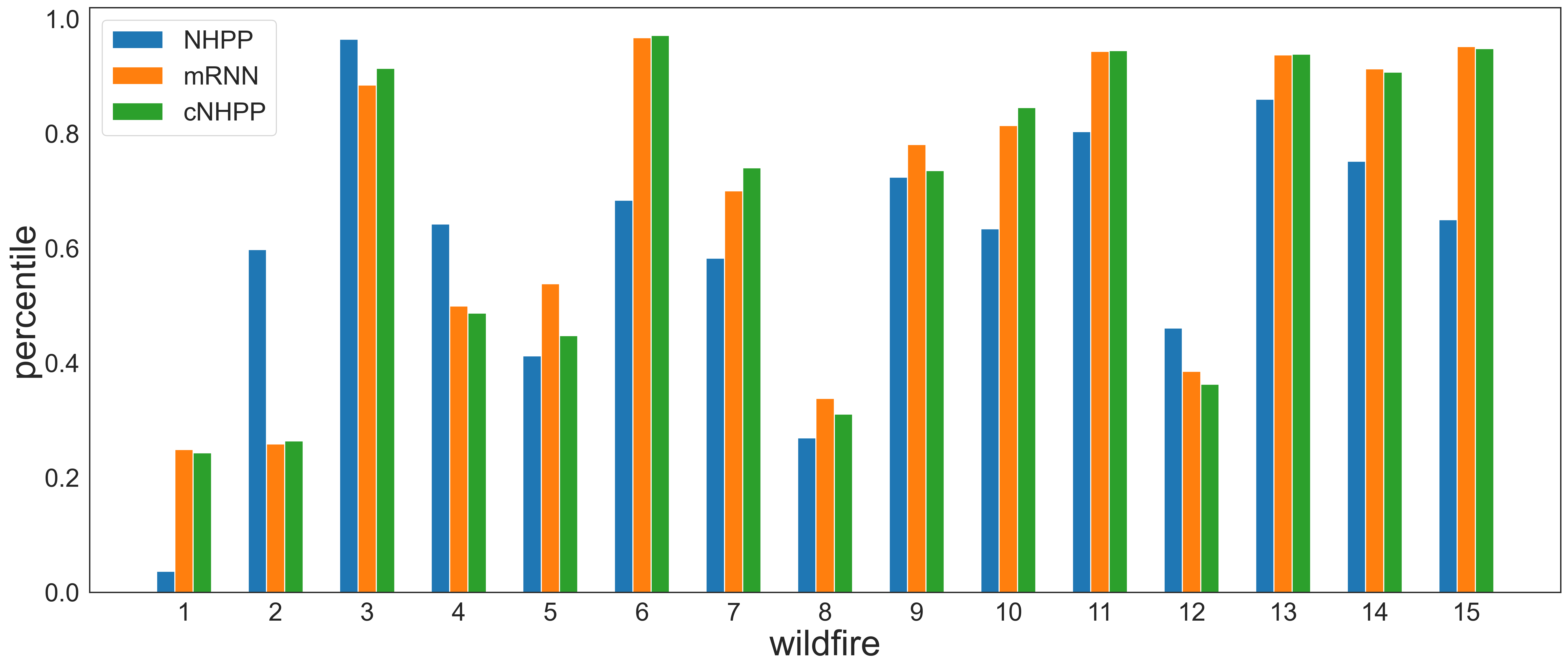}    \caption{The percentiles of the estimated intensities associated with those power lines where fire events occur.}
    \label{fig:percentile}
\end{figure}

\vspace{-8pt}
We further validate the proposed approach in quantifying wildfire risks over the network of power lines. Note that, there are two challenges associated with validating the proposed approach: (i) because only 15 fire incidents are included in the training datasets and the majority of the power line do not experience any fires, traditional measures (such as the mean prediction error, C-index, etc.), become less effective unless we have a much bigger dataset with a much larger number of fire incidents; (ii) power lines with high fire intensity may not always have fire incidents, while power lines with relatively low fire intensity can occasionally catch fire. Hence, we validate the capabilities of the proposed model by utilizing a straightforward but interpretable procedure: for each fire incident, we firstly order the estimated fire intensities from the lowest to the highest for all power transmission lines, and then compute the percentile of the estimated intensity associated with the power line where the wildfire incident occurs. If the model works well, we expect the calculated percentiles associated with those power lines with fire events to be high. The results are shown in Figure \ref{fig:percentile}. It is seen that, among 11 out of the 15 wildfire incidents, the proposed model and its inherited RNN yield higher percentiles than \texttt{NHPP}, suggesting the effectiveness of the proposed approach that accounts for historical cumulative effects and spatial dependency when modeling wildfire risks. 

\vspace{-8pt}
\section{Conclusions}
\vspace{-8pt}
This paper proposed a Convolutional Non-homogeneous Poisson Process (cNHPP) on a linear network. On each network segment, the intensity function is given by the sum of two components. The first component is used to capture the effects of current covariate information, while the second term is used to capture the effects of covariates in the previous time step due to spatial-temporal dependency among neighboring network segments. As a result, the paper showed that the intensity function can be given by the sum of an infinite series, where each term of the series captures the effects of historical covariate information. The paper provided detailed discussions on how the two components are constructed, the computation of the intensity function, the graphical representation of the proposed cNHPP, and how the proposed approach is different from the existing self-exciting process. A natural connection between the proposed method and the Recurrent Neural Network has been drawn. In the application example, we have successfully applied the proposed approach to model and predict the wildfire risks over a network of power transmission lines in California. The model well explained how weather and vegetation variables affect the wildfire risks on power transmission lines and provided some useful insights for mitigating wildfire risks. Comparison studies have been performed to validate the capability of the proposed approach. Computer code is made available at \href{https://github.com/paper-review111/Convolutional-NHPP-Wildfire-Risk-Quantification}{https://github.com/paper-review111/Convolutional-NHPP-Wildfire-Risk-Quantification}.

\begin{center}
{\large\bf SUPPLEMENTARY MATERIAL}
\end{center}

\vspace{-28pt}
\begin{description}
\item Appendices A and B




\end{description}

\bibliographystyle{apalike}
\bibliography{references}

\begin{thebibliography}{}

\bibitem[Baddeley et~al., 2021]{baddeley2021analysing}
Baddeley, A., Nair, G., Rakshit, S., McSwiggan, G., and Davies, T.~M. (2021).
\newblock Analysing point patterns on networks—a review.
\newblock {\em Spatial Statistics}, 42:100435.

\bibitem[Brown et~al., 2021]{brown2021us}
Brown, E.~K., Wang, J., and Feng, Y. (2021).
\newblock $\text{US}$ wildfire potential: A historical view and future
  projection using high-resolution climate data.
\newblock {\em Environmental Research Letters}, 16(3):034060.

\bibitem[CPUC, 2014]{CPUC}
CPUC (2014).
\newblock Decision adopting regulations to reduce the fire hazards associated
  with overhead electric utility facilities and aerial communications
  facilities.
\newblock
  \href{https://docs.cpuc.ca.gov/PublishedDocs/Published/G000/M087/K892/87892306.PDF}{https://docs.cpuc.ca.gov/PublishedDocs/Published/G000/M087/K892/87892306.PDF}.

\bibitem[D’Angelo et~al., 2022]{d2022inhomogeneous}
D’Angelo, N., Adelfio, G., Abbruzzo, A., and Mateu, J. (2022).
\newblock Inhomogeneous spatio-temporal point processes on linear networks for
  visitors’ stops data.
\newblock {\em The Annals of Applied Statistics}, 16(2):791--815.

\bibitem[Fan et~al., 2021]{fan2021selective}
Fan, J., Ma, C., and Zhong, Y. (2021).
\newblock A selective overview of deep learning.
\newblock {\em Statistical science: a review journal of the Institute of
  Mathematical Statistics}, 36(2):264.

\bibitem[Gilardi et~al., 2021]{gilardi2021non}
Gilardi, A., Borgoni, R., and Mateu, J. (2021).
\newblock A non-separable first-order spatio-temporal intensity for events on
  linear networks: an application to ambulance interventions.
\newblock {\em arXiv preprint arXiv:2106.00457}.

\bibitem[Holbrook et~al., 2022]{holbrook2022bayesian}
Holbrook, A.~J., Ji, X., and Suchard, M.~A. (2022).
\newblock Bayesian mitigation of spatial coarsening for a hawkes model applied
  to gunfire, wildfire and viral contagion.
\newblock {\em The Annals of Applied Statistics}, 16(1):573--595.

\bibitem[Hoover and Hanson, 2021]{hoover2021wildfire}
Hoover, K. and Hanson, L.~A. (2021).
\newblock Wildfire statistics.
\newblock Technical report, Congressional Research Service.

\bibitem[Huesca et~al., 2014]{huesca2014modeling}
Huesca, M., Litago, J., Merino-de Miguel, S., Cicuendez-L{\'o}pez-Oca{\~n}a,
  V., and Palacios-Orueta, A. (2014).
\newblock Modeling and forecasting modis-based fire potential index on a pixel
  basis using time series models.
\newblock {\em International Journal of Applied Earth Observation and
  Geoinformation}, 26:363--376.

\bibitem[Katzfuss et~al., 2020]{katzfuss2020ensemble}
Katzfuss, M., Stroud, J.~R., and Wikle, C.~K. (2020).
\newblock Ensemble kalman methods for high-dimensional hierarchical dynamic
  space-time models.
\newblock {\em Journal of the American Statistical Association},
  115(530):866--885.

\bibitem[Kuusela and Stein, 2018]{kuusela2018locally}
Kuusela, M. and Stein, M.~L. (2018).
\newblock Locally stationary spatio-temporal interpolation of argo profiling
  float data.
\newblock {\em Proceedings of the Royal Society A}, 474(2220):20180400.

\bibitem[Mildenberger et~al., 2022]{mildenberger2022effect}
Mildenberger, M., Howe, P.~D., Trachtman, S., Stokes, L.~C., and Lubell, M.
  (2022).
\newblock The effect of public safety power shut-offs on climate change
  attitudes and behavioural intentions.
\newblock {\em Nature Energy}, 7(8):736--743.

\bibitem[Mitchell, 2013]{mitchell2013power}
Mitchell, J.~W. (2013).
\newblock Power line failures and catastrophic wildfires under extreme weather
  conditions.
\newblock {\em Engineering Failure Analysis}, 35:726--735.

\bibitem[MODIS, 2015]{MODIS}
MODIS (2015).
\newblock $\text{MODIS}$ surface reflectance user’s guide collection 6.
\newblock
  \href{https://modis-land.gsfc.nasa.gov/pdf/MOD09_UserGuide_v1.4.pdf}{https://modis-land.gsfc.nasa.gov/pdf/MOD09\_UserGuide\_v1.4.pdf}.

\bibitem[Mohler et~al., 2011]{mohler2011self}
Mohler, G.~O., Short, M.~B., Brantingham, P.~J., Schoenberg, F.~P., and Tita,
  G.~E. (2011).
\newblock Self-exciting point process modeling of crime.
\newblock {\em Journal of the American Statistical Association},
  106(493):100--108.

\bibitem[Nazaripouya, 2020]{nazaripouya2020power}
Nazaripouya, H. (2020).
\newblock Power grid resilience under wildfire: A review on challenges and
  solutions.
\newblock In {\em 2020 IEEE Power \& Energy Society General Meeting (PESGM)},
  pages 1--5. IEEE.

\bibitem[NFDRS, 2002]{NFDRS}
NFDRS (2002).
\newblock Gaining an understanding of the national fire danger rating system.
\newblock
  \href{https://www.nwcg.gov/sites/default/files/products/pms932.pdf}{https://www.nwcg.gov/sites/default/files/products/pms932.pdf}.

\bibitem[Opitz et~al., 2020]{opitz2020point}
Opitz, T., Bonneu, F., and Gabriel, E. (2020).
\newblock Point-process based bayesian modeling of space--time structures of
  forest fire occurrences in mediterranean france.
\newblock {\em Spatial Statistics}, 40:100429.

\bibitem[Peng et~al., 2005]{peng2005space}
Peng, R.~D., Schoenberg, F.~P., and Woods, J.~A. (2005).
\newblock A space--time conditional intensity model for evaluating a wildfire
  hazard index.
\newblock {\em Journal of the American Statistical Association},
  100(469):26--35.

\bibitem[Rhodes and Roald, 2022]{rhodes2022co}
Rhodes, N. and Roald, L. (2022).
\newblock Co-optimization of power line shutoff and restoration for electric
  grids under high wildfire ignition risk.
\newblock {\em arXiv preprint arXiv:2204.02507}.

\bibitem[Schulze et~al., 2020]{schulze2020wildfire}
Schulze, S.~S., Fischer, E.~C., Hamideh, S., and Mahmoud, H. (2020).
\newblock Wildfire impacts on schools and hospitals following the 2018
  california camp fire.
\newblock {\em Natural Hazards}, 104(1):901--925.

\bibitem[SDG\&E, 2021]{SDGE}
SDG\&E (2021).
\newblock San diego gas \& electric company 2020‐2022 wildfire mitigation
  plan update.
\newblock
  \href{https://www.sdge.com/sites/default/files/regulatory/SDG\%26E\%202021\%20WMP\%20Update\%2002-05-2021.pdf}{https://www.sdge.com/sites/default/files/regulatory/SDG\%26E\%202021\%20WMP\\\%20Update\%2002-05-2021.pdf}.

\bibitem[Serra et~al., 2014]{serra2014spatio}
Serra, L., Saez, M., Mateu, J., Varga, D., Juan, P., D{\'\i}az-{\'A}valos, C.,
  and Rue, H. (2014).
\newblock Spatio-temporal log-gaussian cox processes for modelling wildfire
  occurrence: the case of catalonia, 1994--2008.
\newblock {\em Environmental and Ecological Statistics}, 21(3):531--563.

\bibitem[Stein, 2005]{stein2005space}
Stein, M.~L. (2005).
\newblock Space--time covariance functions.
\newblock {\em Journal of the American Statistical Association},
  100(469):310--321.

\bibitem[Taylor et~al., 2013]{taylor2013wildfire}
Taylor, S.~W., Woolford, D.~G., Dean, C., and Martell, D.~L. (2013).
\newblock Wildfire prediction to inform fire management: statistical science
  challenges.
\newblock {\em Statistical Science}, 28(4):586--615.

\bibitem[Trilles et~al., 2013]{trilles2013integration}
Trilles, S., Juan, P., Diaz, L., Arago, P., and Huerta, J. (2013).
\newblock Integration of environmental models in spatial data infrastructures:
  a use case in wildfire risk prediction.
\newblock {\em IEEE Journal of Selected Topics in Applied Earth Observations
  and Remote Sensing}, 6(1):128--138.

\bibitem[Uppala and Handcock, 2020]{uppala2020modeling}
Uppala, M. and Handcock, M.~S. (2020).
\newblock Modeling wildfire ignition origins in southern california using
  linear network point processes.
\newblock {\em The Annals of Applied Statistics}, 14(1):339--356.

\bibitem[Vazquez et~al., 2022]{vazquez2022wildfire}
Vazquez, D. A.~Z., Qiu, F., Fan, N., and Sharp, K. (2022).
\newblock Wildfire mitigation plans in power systems: A literature review.
\newblock {\em IEEE Transactions on Power Systems}.

\bibitem[Verbesselt et~al., 2006]{verbesselt2006evaluating}
Verbesselt, J., Jonsson, P., Lhermitte, S., Van~Aardt, J., and Coppin, P.
  (2006).
\newblock Evaluating satellite and climate data-derived indices as fire risk
  indicators in savanna ecosystems.
\newblock {\em IEEE Transactions on Geoscience and Remote Sensing},
  44(6):1622--1632.

\bibitem[Xu and Schoenberg, 2011]{xu2011point}
Xu, H. and Schoenberg, F.~P. (2011).
\newblock Point process modeling of wildfire hazard in los angeles county,
  california.
\newblock {\em The Annals of Applied Statistics}, 5(2A):684--704.

\bibitem[Zhu et~al., 2022]{zhu2022spatio}
Zhu, S., Yao, R., Xie, Y., Qiu, F., Qiu, Y.~L., and Wu, X. (2022).
\newblock Quantifying grid resilience against extreme weather using large-scale
  customer power outage data.
\newblock {\em arXiv preprint arXiv:2109.09711}.

\end{thebibliography}
\end{document}